\pdfoutput=1
 \documentclass[twocolumn,amsmath,amssymb,letter, preprintnumbers]{revtex4}
\makeatletter
\newif\if@restonecol
\makeatother

\usepackage[boxed]{algorithm2e} 
\usepackage{graphicx}
\usepackage{dcolumn}
\usepackage{bm}

\begin{document}

\title{Automatic Metadata Generation using Associative Networks\footnote{Rodriguez M.A., Bollen, J., Van de Sompel, H., ``Automatic Metadata Generation using Associative Networks",  ACM Transactions on Information Systems, volume 27, number 2, pages 1-20, doi:10.1145/1462198.1462199,Ê ISSN: 1046-8188, ACM Press, February 2009.}}

\author{Marko A. Rodriguez}
\email{marko@lanl.gov}
\affiliation{Digital Library Research and Prototyping Team \\
		Los Alamos National Laboratory \\
		Los Alamos, New Mexico 87545 }

\author{Johan Bollen}
\email{jbollen@lanl.gov}
\affiliation{Digital Library Research and Prototyping Team \\
		Los Alamos National Laboratory \\
		Los Alamos, New Mexico 87545 }
		
\author{Herbert Van de Sompel}
\email{herbertv@lanl.gov}
\affiliation{Digital Library Research and Prototyping Team \\
		Los Alamos National Laboratory \\
		Los Alamos, New Mexico 87545 }

\preprint{LA-UR-06-3445}

\begin{abstract}
In spite of its tremendous value, metadata is generally sparse and incomplete, thereby hampering the effectiveness of digital information services. Many of the existing mechanisms for the automated creation of metadata rely primarily on content analysis which can be costly and inefficient. The automatic metadata generation system proposed in this article leverages resource relationships generated from existing metadata as a medium for propagation from metadata-rich to metadata-poor resources. Because of its independence from content analysis, it can be applied to a wide variety of resource media types and is shown to be computationally inexpensive. The proposed method operates through two distinct phases. Occurrence and co-occurrence algorithms first generate an associative network of repository resources leveraging existing repository metadata. Second, using the associative network as a substrate, metadata associated with metadata-rich resources is propagated to metadata-poor resources by means of a discrete-form spreading activation algorithm. This article discusses the general framework for building associative networks, an algorithm for disseminating metadata through such networks, and the results of an experiment and validation of the proposed method using a standard bibliographic dataset.
\end{abstract}

\maketitle

\section{Introduction}

Resource metadata plays a pivotal role in the functionality and interoperability of digital information repositories. However, in spite of its value, high quality metadata is difficult to come by \cite{duval:meta2002}. \cite{dc:ward2003} demonstrates that although as many as 15 possible metadata properties can theoretically be included in the widely used Dublin Core standard\footnote{The Dublin Core 1.1 specification is available at: http://dublincore.org/documents/1999/07/02/dces/}, few are frequently used in collections whose metadata are generally created by the author's themselves\footnote{The most frequently used are \textit{creator}, \textit{identifier}, \textit{title}, \textit{date}, and \textit{type}.}. The problem of poor and incomplete metadata is expected to worsen as repositories are applied to materials collected beyond the traditional, centralized methods of publication and start to obtain data from web pages, blogs, personal multimedia collections, and collaborative tagging environments.\\

Metadata is a costly resource to create, maintain, and/or recover manually. There has therefore been significant research on automated metadata generation (e.g.~by extracting metadata from the content of resources). Natural language processing \cite{meta:yang2005} and document image analysis techniques \cite{meta:guirida2000,sebastiani02machine,meta:han2003,meta:mao2004} may extract keywords, subject categories, author, and citations (e.g.~CiteSeer\footnote{CiteSeer available at: http://citeseer.ist.psu.edu/}) from manuscripts. Furthermore, in \cite{meta:greenburg2004}, two metadata generators are demonstrated that successfully harvest and extract metadata from existing resource source and content. Such content-based techniques are much less efficient for multimedia resources, e.g.  video, music, images, and datasets. Reliable content analysis for such data is still an active research area and existing methods generally yield little content-related metadata. In addition, content-based approaches can be prohibitively expensive in computational terms \cite{kuwano:meta2004}.\\

For the reasons outlined above, methods for the generation of metadata that do not rely on resource content have generated considerable interest. The recent growth in applications of ``folksonomies" (i.e.~community-based ``tagging" \cite{tag:mathes2004,tagging:hub2006}), has been, to some extent, inspired by the shortcomings of existing metadata generation methods. Unfortunately, human tagging only works well in situations where the number of participants greatly exceeds the number of resources to be tagged and where there is no requirement for controlled vocabularies or standardized metadata formats.\\

In this article, we propose a system for automated metadata generation that starts from a common scenario: a heterogeneous repository contains resources for which varying degrees of metadata are available. Some resources have been imbued with rich, vetted metadata, whereas others have not. However, if it can be assumed that  resources that are ``similar" (e.g.~similar in publication venue, authorship, date, citations, etc.) are more likely to have shared metadata, then the problem of metadata generation can be reformulated as one of extrapolating metadata from metadata-rich to related, but metadata-poor resources. This article's experiment focuses on identifying which aspects of metadata similarity are best used to extrapolate resource metadata in a bibliographic dataset.\\

As a case in point, \cite{levera:naaman2005} describes a method to support the annotation of personal photograph collections. Once a user has annotated a photograph its metadata is automatically transferred to photographs taken at similar times and locations. For example, a user photographs a group of friends at 3:45PM. Another photograph is made at 3:47PM.  Since the second photograph was taken only two minutes after the first, it is likely that it depicts a similar scene. The system therefore transfers metadata from photograph 1 to photograph 2. Similarly, \cite{metaprop:prime2005} proposes a method of web page metadata propagation using co-citation networks. The general idea is that if two web pages cite other web pages in common, then the probability that they share similar metadata is higher. The user can later correct and augment any transferred metadata.\\

The mentioned systems are strongly related to collaborative filtering \cite{collab:herlocker2006}. Collaborative filtering systems are commonly employed in online retail systems to recommend items of interest to individual users. Using the principle that similar users are more likely to appreciate similar items, users are recommended items that are missing from their profiles but occur in the profiles of similar users. The collaborative filtering process can thus be regarded as an instance of metadata propagation. If users are considered resources and their profiles are considered ``resource metadata", it can be said that collaborative filtering systems ``recommend" metadata from one resource to another based on resource similarity.\\

A generalization of the above metadata propagation systems can be made in terms of the following elements:

\begin{enumerate}
	\item A mechanism to generate resource relations, i.e. assess their similarity.
	\item The determination of a metadata-rich subset of the repository's collection that can serve as a reference set.
	\item A means of propagating metadata from the metadata-rich reference set to a metadata-poor subset of the collection using the established resource relations as a substrate.
\end{enumerate}

Such systems for the generation of metadata can be said to operate on a ``Robin Hood" principle; they take from metadata-rich resources and give to metadata-poor resources, with the exception that metadata is not a zero-sum resource. This mode of operation has a number of desirable properties. First, it reduces the need for the costly generation of metadata; metadata is automatically extrapolated from an existing metadata-rich reference collection to a metadata-poor subset. Second, resource relations can be defined independent of content and metadata extrapolation can thus be implemented for wide range of heterogeneous resources, e.g.~audio, video, and images.\\

\begin{figure}[h!]
	\centering
		\includegraphics[width=0.5\textwidth]{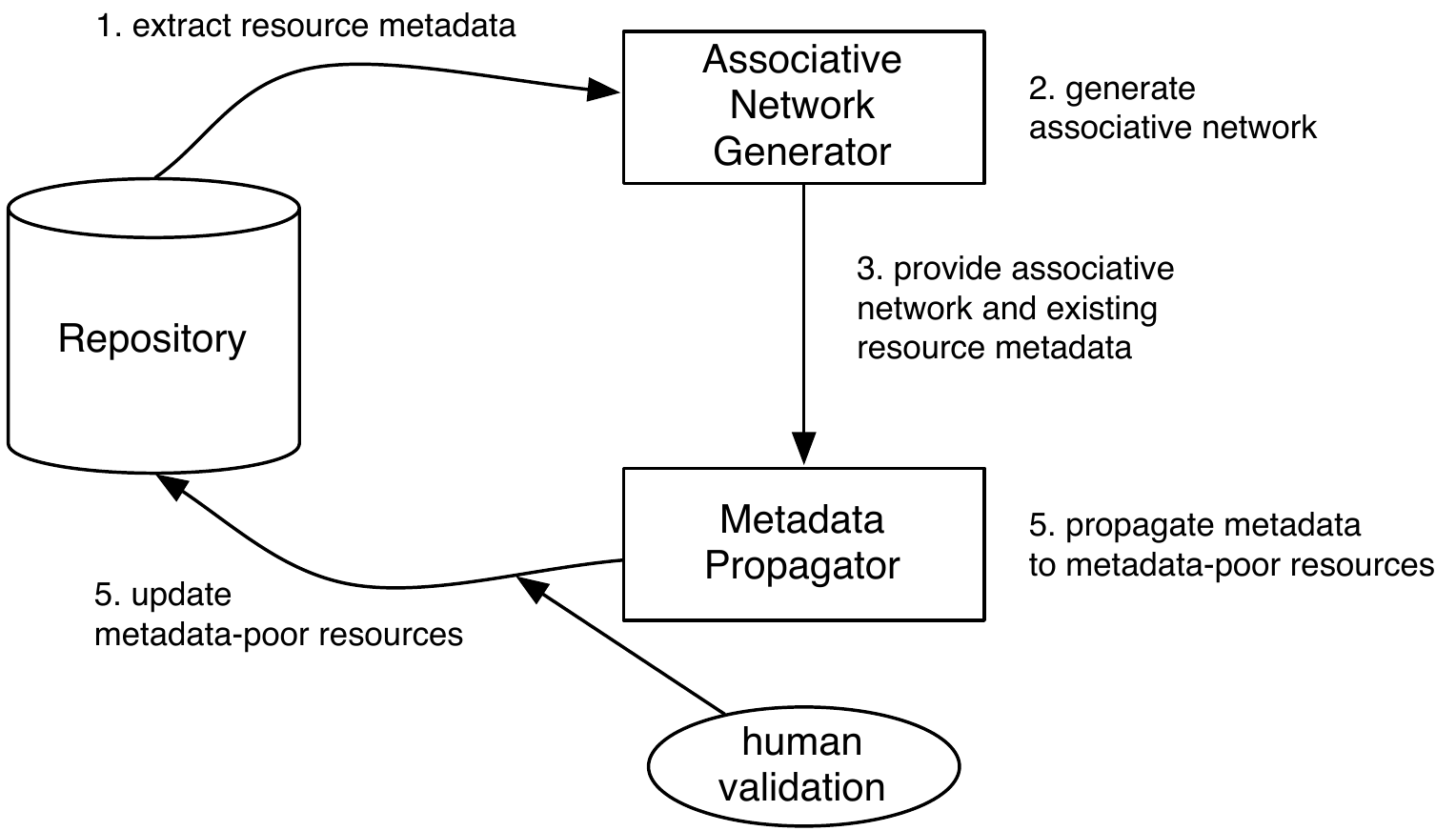}
	\caption{System outline}
	\label{fig:outline}
\end{figure}

This article outlines a proposal for a metadata propagation system designed for scholarly repositories that takes advantage of the multiple means by which two resources can be related (e.g.~co-citation, citation, co-author, co-keyword, etc.). Figure \ref{fig:outline} presents the outline of the proposed system's components and processing stages. First, resource metadata is extracted from the collection of a repository. Second, an associative multi-relational network (i.e.~a directed labeled graph) of resource relations is derived from a subset of the available metadata. Third, a metadata-rich subset of the collection is selected to serve as a reference data set. Fourth, and finally, metadata is propagated (i.e.~extrapolated) from the metadata-rich reference set to all other metadata-poor resources over the associative network of resources after which the repository is updated. Human validation can vet the results of the metadata extrapolation before insertion into the repository occurs.\\

It is important to emphasize that this system requires the existence of some preliminary metadata both for the construction of resource relations and for metadata propagation. Furthermore, the quality or accuracy of the preliminary metadata is important in ensuring successful results (i.e.~to avoid a ``garbage in, garbage out" scenario). However, the metadata being propagated can be different from the metadata used to generate resource relations. For instance, in the manuscript domain, the propagation of keyword metadata may be most efficient along resource relations derived from citation metadata. Therefore, two aspects affect the efficiency of metadata propagation: the type of resource relations and the algorithm used to propagate metadata. It is important to note that no new metadata values are created in model proposed in this article. While it is important for resources to maintain metadata, this method only propagates pre-existing metadata values and thus, does not increase the discriminatory aspects that metadata should and generally provides. While like resources should have similar metadata, variations should also exist to make sure that a resource's metadata accentuates the unique characteristics of the resource.\\

This paper will first discuss two algorithms to define sets of resource relations and represent these relations in terms of associative networks. It will then formally define a metadata propagation algorithm which can operate on the basis of the generated resource relations. Finally, the proposed metadata generation system is validated using a modified version of the KDD Cup 2003 High-Energy Physics bibliographic dataset (hep-th 2003)\footnote{hep-th 2003 available at: http://www.cs.cornell.edu/projects/kddcup/datasets.html}. While it is theoretically possible for this method to work on other resource types (e.g.~video, audio, etc.) as it doesn't require an analysis of the content of the resources, only their metadata; it is only speculated that the results of such a method would be viable in these other, non-tested, domains.

\section{Constructing an Associative Network of Repository Resources}

An associative network is a network that connects resources according to some measure of similarity. An associative network is represented by the data structure $G =(N,E,W)$ where $N$ is the set of resources, $E \subseteq N \times N$ the set of directed relationships between resources, and $W$ is the set of weight values for all edges such that $|W| = |E|$. Any edge $e_{i,j,\mu}$ with corresponding weight $w_{i,j,\mu}$ expresses that there exists a directed weighted relationship constructed using properties of type $\mu$ from resource $n_i$ to resource $n_j$. The explicit representation of $\mu$ is necessary because an associative network can be constructed according to different properties (i.e.~authorship, citations, keywords, etc.). As will be demonstrated, certain network $\mu$ relationships are better (in terms of precision and recall) at propagating certain property types than others.\\

The remainder of this section will describe two associative network construction algorithms. One is based on {\it occurrence} metadata where a resource is considered similar to another if there is a direct reference from one resource to the other (e.g.~a direct citation). The other algorithm is based on {\it co-occurrence} metadata and thus, considers two resources to be similar if they share similar metadata. That is, two resources are deemed similar if the same metadata values occur in both their properties (i.e.~same authors, same keywords, same publication venue, etc.). Depending on how the repository represents its metdata some property types will be direct reference properties and others will have to be infered through indirect, co-occurence algorithms.\\

\subsection{Occurrence Associative Networks}

An associative network can be constructed if direct references connect one resource to another. The World Wide Web, for instance, is an associative network based on occurrence data because a web-page makes a direct reference to another web-page via a hyper-link (i.e.~the \textit{href} HTML tag). For manuscript resources, occurrence information usually exists in citations. For instance, if resource $n_i$ references (i.e.~cites) resource $n_j$ then their exists an edge $e_{i,j,\mathrm{cite}}$. One potential algorithm for determining the edge weight is to first determine how many other citations resource $n_i$ currently maintains.  That is, if resource $n_i$ also cites $50$ other resources then resource $n_i$ is $\frac{1}{50}$ as similar to $n_j$, $w_{i,j,\mathrm{cite}} = \frac{1}{50}$. Similarly, if resource $n_i$ only cites resource $n_j$ then the strength of tie to resource $n_j$ is greater, $w_{i,j,\mathrm{cite}} = 1.0$.  The general equation is defined by Eq. \ref{eq:cite} where the function $\mathrm{meta}(n_i,\mathrm{cite})$ returns the set of all citations for resource $n_i$.  This equation only holds if resource $n_j \in \mathrm{meta}(n_i,\mathrm{cite})$. Eq. \ref{eq:cite} makes use of the $\mu$ notation in order to generalize the equation for use with any direct reference property types.\\

	\begin{equation*}
		\label{eq:cite}
		w_{i,j,\mathrm{\mu}} = \frac{1}{\left|\mathrm{meta}(n_i,\mathrm{\mu})\right|} \; : \; n_j \in \mathrm{meta}(n_i,\mathrm{\mu})
	\end{equation*}
  
The running time of the algorithm to construct an associative network based on direct, occurrence property types is $O(N)$ since each resource must be checked once and only once for direct reference to other resources.\\

\subsection{Co-occurrence Associative Networks}

Co-occurrence networks are created when resources share the same metadata property values. For instance, if two resources share the same keyword, author, or citation values then there exists some degree of similarity. For a co-occurrence network the edge weight for any two resources, $w_{i,j,\mathrm{co}\mu}$ and $w_{j,i,\mathrm{co}\mu}$, is a function of the amount of metadata properties of type $\mu$ that $n_i$ and $n_j$ share in common. A specific example of this could be a co-keyword associative network created when two resources have similar keywords. For example, suppose the resource nodes $n_i$ and $n_j$ have the following list of keyword properties presented in Table \ref{tab:cokey}.\\

\begin{table}[h!]
	\begin{footnotesize}
	\begin{center}
		\begin{tabular}{|c|c|c|c|}
		\hline
		resource & keyword-1 & keyword-2 & keyword-3\\
		\hline
		$n_i$ & \textbf{repository} & \textbf{metadata} & particle \\
		$n_j$ & images & \textbf{repository} & \textbf{metadata} \\
		\hline
		\end{tabular}
	\end{center}
	\end{footnotesize}
	\caption{\label{tab:cokey} Keyword metadata for resources $n_i$ and $n_j$}
\end{table}

In Table \ref{tab:cokey}, resource $n_i$ and $n_j$ share two keywords in common, namely {\it repository} and {\it metadata}. The edge weight between these two resources is a function of the amount of keywords they share in common, Eq. \ref{eq:cokey1}, and the size of the keyword count of both resources. Therefore, according to Eq. \ref{eq:cokey2}, the edges connecting resource $n_i$ to $n_j$ and $n_j$ to $n_i$ have a weight of $w_{n_i,n_j,\mathrm{cokey}} = w_{n_j,n_i,\mathrm{cokey}} = 0.5$.\\

	\begin{equation*}
		\label{eq:cokey1}
		\mathrm{co}(n_i,n_j,\mu) = \mathrm{meta}(n_i,\mu) \cap \mathrm{meta}(n_j,\mu)
	\end{equation*}
	
so that\\
	\begin{footnotesize}
	\begin{equation*}
		\label{eq:cokey2}
		w_{i,j,\mathrm{co}\mu} = w_{j,i,\mathrm{co}\mu} = \frac{|\mathrm{co}(n_i,n_j,\mu)|}{[|\mathrm{meta}(n_i,\mu)| + |\mathrm{meta}(n_j,\mu)|] - |\mathrm{co}(n_i,n_j,\mu)|}
	\end{equation*}
	\end{footnotesize}

Notice that the co-occurrence algorithm in Eq. \ref{eq:cokey2} returns a $\mathrm{co}\mu$ representation.  This means for keyword properties, the returned weight is a co-keyword similarity weight. Similarly, for authorship metadata, the returned weight is a co-authorship weight. The running time of the algorithm to construct a co-occurrence network is $O(\frac{N^2-N}{2})$ since each resource's $\mu$-properties must be checked against every other resource's $\mu$-properties ($N^2$), except itself ($-N$), once and only once ($\frac{1}{2}$).\\

\section{Metadata Propagation Algorithm}

Reconstructing the metadata for a metadata-poor collection of resources is dependent not only on the associative network data structure, but also upon the use of a metadata propagation algorithm. The algorithm chosen is a derivative of the particle-swarm algorithm \cite{socialgrammar:rodriguez2007}. Particle-swarm algorithms are a discrete form of the spreading activation algorithms \cite{spread:collins1975,inform:cohen1987,applic:crestani1997,search:crestani2000,applyi:huang2004}. Because particles are indivisible entities, it is easy to represent metadata properties as being encapsulated inside a particle. These metadata particles are then propagated over the edges of the associative network. Upon reaching a resource node that is missing a particular property type, the particle recommends its property value to the visited resource. This section will formally describe the metadata propagation algorithm before discussing the results of an experiement using a bibliographic dataset.\\

Every resource node in an associative network is supplied with a single particle, $p_i \in P$, such that $|P| = |N|$. The particle $p_i$ encapsulates all the metadata properties of a particular resource $n_i$. Therefore, $\mathrm{meta}(n_i, \mu) \equiv \mathrm{meta}(p_i, \mu)$ for all $\mu$. Particle $p_i$ has a reference to its current node $c_i \in N$ such that at $t=0$, $c_i=n_i$. The particle $p_i$ begins its journey ($t=0$) at its home node, $n_i$, and traverses an outgoing edge of $n_i$. Particle edge traversal is a stochastic process that requires the outgoing edge weights of each node to form a probability distribution. Therefore, the set of outgoing edge weights of relation type $\mu$ for $n_i$, $\mathrm{out}(n_i,\mu)$, must be normalized as represented in Eq. \ref{eq:normedges1} and Eq. \ref{eq:normedges2}. Unlike Eq. \ref{eq:cokey2}, for co-occurrence edges, these equations do not guarantee that $w_{i,j,\mu} = w_{j,i,\mu}$.\\

	\begin{equation*}
		\label{eq:normedges1}
			w_{i,j,\mu} = \frac{w_{i,j,\mu}}{\sum_{\forall k \in e_{i,k,\mu}} w_{i,k,\mu}}
	\end{equation*}

such that\\

	\begin{equation*}
		\label{eq:normedges2}
			\sum_{\forall j \in e_{i,j,\mu}} w_{i,j,\mu} = 1.0
	\end{equation*}

The function $\theta(\mathrm{out}(n_i,\mu))$ is defined such that it takes a set of outgoing edges of relation type $\mu$ of node $n_i$ and returns a single node $n_j$ based upon the outgoing edge weight probability distribution, where $e_{i,j,\mu} \in \mathrm{out}(n_i,\mu)$. This is how a particle traverses an associative network.\\

The particle $p_i$ also has an associated energy value $\epsilon_i \in [0,1]$. Each time an edge is traversed, the particle $p_i$ decays its energy content, $\epsilon_i$, according to a global decay value, $\delta \in [0,1]$.  Particle energy decay over discrete time $t$ is represented in Eq. \ref{eq:pathdecay}. The rational for decay is based on the intuition that the metadata property values of a particular particle become less relevant the further the particle travels away from its source node ($c_i$ at $t=0$). Therefore, the further a particle travels in the network, the more that particle's energy value (or recommendation influence), $\epsilon$, is decayed.\\

	\begin{equation*}
		\label{eq:pathdecay}
			\epsilon_i(t+1) = (1 - \delta)\epsilon_i(t)
	\end{equation*}

\begin{figure}[h!]
	\centering
		\includegraphics[width=0.5\textwidth]{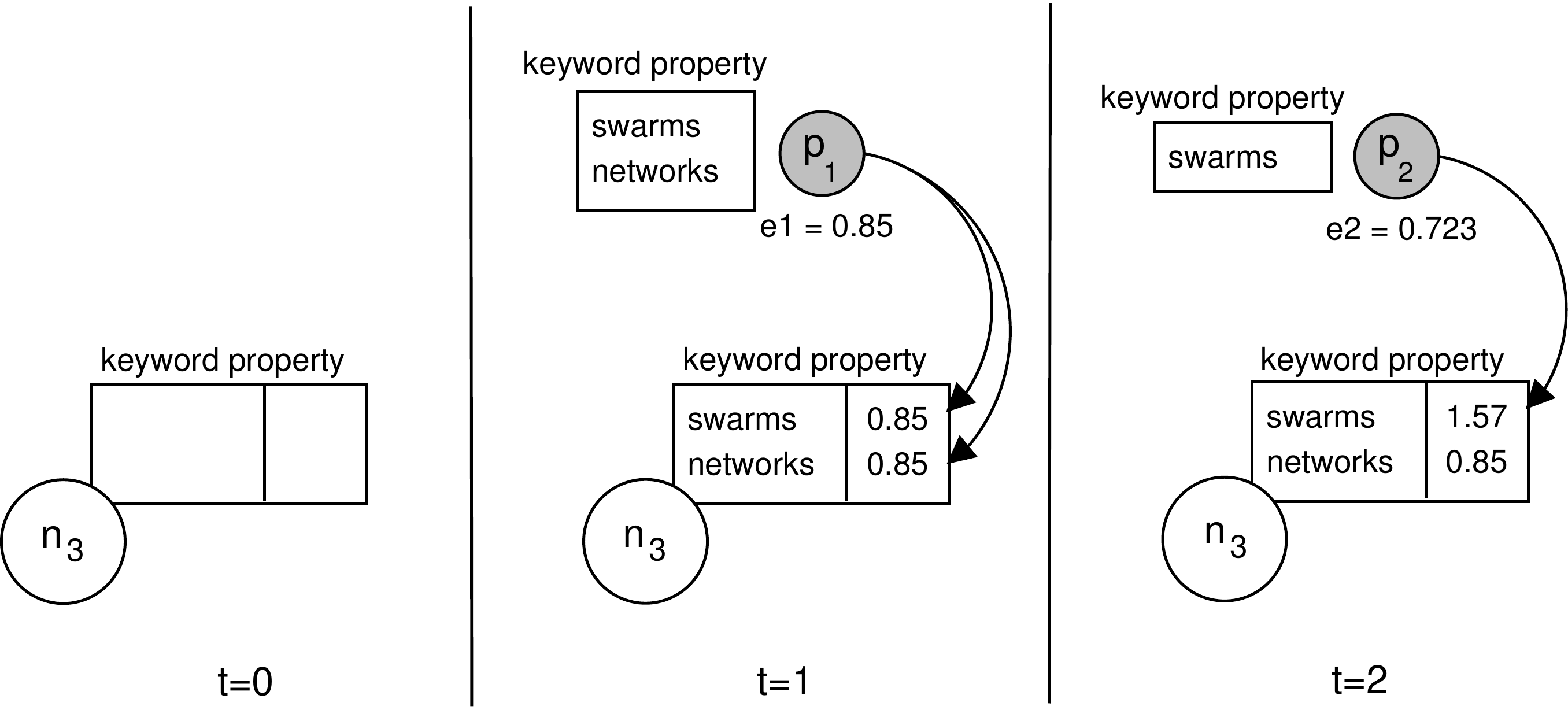}
	\caption{Particles recommending metadata information to a metadata-poor node}
	\label{fig:recommend}
\end{figure}

The energy value of a particle defines how much recommendation influence a particle's metadata property values has on a visited metadata-poor node.  Each time a particle traverses a node with missing metadata properties, it not only recommends its metadata property values to that node, but also increments the appropriate property value with its current energy value $\epsilon_i$. In Figure \ref{fig:recommend}, at $t=0$, before the propagation algorithm has been executed, resource $n_3$ has no keyword values.  Therefore, when particle $p_1$ reaches $n_3$ at $t=1$, particle $p_1$ recommends its keyword property values (keyword=\{swarm, algorithms\}) to node $n_3$ with an influence of $\epsilon_1=0.85$. At $t=2$, particle $p_2$, with $\epsilon_2=0.723$, recommends its keyword property (keyword=\{swarm\}) to node $n_3$. Notice that the recommendation of the keyword property value `swarm' is reinforced each time that property value is presented to $n_3$.\\

The function of a single particle, $p_i$, at a particular node, $n_j$, is represented in pseudo-code in Algorithm \ref{alg:recommendMeta} where $\mathrm{rec}(n_j,\mu)$ returns the set of previous property values to $n_j$ for a property of type $\mu$.\\

  \restylealgo{boxed}
  \linesnumbered
  \begin{algorithm}[h!]
  \begin{footnotesize}
  \Indp
   \KwIn{$\mathrm{recommendMeta}(n_j,p_i)$}
	 		\# $p_i$ updates the metadata of $n_j$ for all property types\;
	  	\ForEach{$(\mu\mathrm{-property})$}{
		 \Indp   \# first ensure that $n_j$ is metadata-poor at the particular $\mu$-property\;
		    \If{$(|\mathrm{meta}(n_j, \mu)| == 0)$}{
			 \Indp     \# update the metadata-poor node's $\mu$-property with the $\mu$ property value of $p_i$\;
			   \ForEach{$(x \in \mathrm{meta}(p_i, \mu))$}{
				 \Indp  	$\mathrm{found} = false$\;
				  	\# if property value already exists, increment its energy value with $e_i$\;
				   	\ForEach{$(y \in \mathrm{rec}(n_j, \mu))$}{
				 \Indp	  	\If{$(x == \mathrm{value}(y))$}{
				 \Indp	  		$\mathrm{energy}(y) = \mathrm{energy}(y) + \epsilon_i$\;
					  		$\mathrm{found} = true$\;
					  }
					 }
					 \# if no recommended value exists, add to $n_j$'s recommendations\;
					 \If{$(!\mathrm{found})$}{
				 \Indp	  $\mathrm{addRec}(x,\epsilon_i) = x$\;
					}
				}
			}
		}
	\caption{\label{alg:recommendMeta} Particle $p_i$ recommending metadata properties values to $n_j$}
  \end{footnotesize}
  \end{algorithm}
	
If Algorithm \ref{alg:recommendMeta} is called $\mathrm{recommendMeta}(n_j, p_i)$ then the full particle propagation algorithm can be described by the pseudo-code in Algorithm \ref{alg:propagate}. The process of moving metadata particles through the associative network and recommending metadata-poor nodes metadata property values continues until some desired $t$ is reached or all particle energy in the network has decayed to $0.0$, $\sum_{\forall i} \epsilon_i \cong 0.0$.\\

  \restylealgo{boxed}
  \linesnumbered
  \begin{algorithm}[h!]
  \begin{footnotesize}
  \Indp
	  \KwIn{$\mathrm{propagate}(\mu)$}
	  \# $\delta$ is a global energy decay value \;
	  $\delta = 0.15$ \;
		\# create a particle for each node \;
	  \ForEach{$(n_i \in N)$}{
	 \Indp		$\mathrm{meta}(p_i,\mu) = \mathrm{meta}(n_i,\mu) \;:\; \forall \mu$ \;
			$\epsilon_i = 1.0$ \;
			$c_i = n_i$ \;
		 }
		 \# propagate metadata particles throughout $\mu$ network \;
		 $t = 0$ \;
		 \While{$(\sum_{\forall p_i \in P} \epsilon_i > 0.0 \;\; \&\& \;\; t < \mathrm{maxSteps})$}{
	 \Indp		 \ForEach{$(p_i \in P)$}{
	 \Indp			\# if $c_i$ has no outgoing edges, freeze the particle \;
					\If{$(|\mathrm{out}(c_i,\mu)| > 0)$}{
	 \Indp				$c_i = \theta(\mathrm{out}(c_i,\mu))$ \;
					$\epsilon_i = \epsilon_i * (1-\delta)$ \;
					\# do not recommend metadata to the particle's home node \;
					\If{$(c_i \; \mathrm{!=} \; n_i)$}{
	 \Indp					$\mathrm{recommendMeta}(c_i, p_i)$ \;	
					}
				}
			 }
			 $t = t + 1$ \;
	   }
	\caption{\label{alg:propagate} Propagating metadata particles through an associative network of type $\mu$}
  \end{footnotesize}
  \end{algorithm}

\section{An Experiment using the 2003 HEP-TH Bibliographic Dataset}

This section will present the results of the proposed metadata generation system when attempting to reconstruct an artificially atrophied bibliographic dataset. By artificially reducing the amount of metadata in the full bibliographic dataset, it is possible to simulate a metadata-poor environment and at the same time still be able to validate the results of the metadata propagation algorithm. The section is outlined as follows.  First, the dataset used for this experiment is described. Second, a short review of the validation metrics (precision, recall, and F-score) is presented. Third, the various system parameters are discussed. Finally, the results of the experiment are presented as a validation of the systems use for manuscript-based digital library repositories. Further research into other domains besides manuscripts will demonstrate the validity of this method for other resource types.\\

The dataset used to validate the proposed system is a modified version of the hep-th 2003 bibliographic dataset for high energy physics and theory \cite{mcgovern03exploiting}.\footnote{The details of how the hep-th dataset was created can be found in the disseminated README (http://kdl.cs.umass.edu/data/hepth/hepth-README.txt) of the data set and a few specifics are quoted here: ``Object and link properties such as title, authors, journal (if published), and various dates were extracted from the abstract files with additional date information coming from the slacdates citation tarball.  Because institutions were often not presented in a consistent format, the email domain of the submitter (if available) was used as a surrogate for institution.  Because many authors had no associated email address, domain information is not available for all authors.  Consolidation was performed on journal names, domains, and author names.  A nominal amount of hand-cleaning to correct spelling or formatting problems was also performed."}  A modified version of the hep-th dataset, as used in \cite{unsupervised:lin2004}, is represented as a semantic network containing manuscripts (29,014), authors (12,755), journals (267), organizations (963), keywords (40), and publication date in year/season pairs (60).  These nodes are then connected according to the following semantics:\\

\begin{itemize}
	\item writes($a$,$m$): author $a$ wrote manuscript $m$
	\item date\_published($m$,$d$): manuscript $m$ was published on date $d$
	\item organization\_of($a$,$o$): author $a$ works for organization $o$
	\item published\_in($m$,$j$): manuscript $m$ was published in journal $j$
	\item cites($m_x$,$m_y$): manuscript $m_x$ cites manuscript $m_y$
	\item keyword\_of($m$,$k$): manuscript $m$ has keyword $k$ 
\end{itemize}

For the purposes of this experiment, the semantic network from \cite{unsupervised:lin2004} was transformed into a list of manuscripts and their associated metadata property name/value pairs. These manuscript properites include: authors, date of publication, citations, keywords, publishing journal, and organizations. Of the 29,014 manuscript nodes, different occurrence and co-occurrence algorithms were used to construct the following associative networks:\\

\begin{enumerate}
	\item citation: manuscript $m_i$ maintains an edge to manuscript $m_j$ if $m_i$ cites $m_j$ (27,240 edges)
	\item co-author: manuscripts maintain an edge if they share authors (724,406 edges)
	\item co-citation: manuscripts maintain an edge if they share citations (23,089,616 edges)
	\item co-keyword: manuscripts maintain an edge if they share keywords (12,418,172 edges)
	\item co-organization: manuscript maintain an edge if they share organizations (33,947,083 edges)
\end{enumerate}

Though not explored empirically, it is worth noting that link prediction algorithms can be employed to resolve issues relating to edge sparsity in the network. In particular, the methods proposed in \cite{linkpr:liben2003} and \cite{linkpr:hunag2005} are such algorithms.\\

\subsection{A Review of Precision, Recall, and F-Score}

The results of the metadata generation experiment are evaluated according to the F-score measure so therefore, it is important to provide a quick review of {\it precision}, {\it recall}, and {\it F-score} within the framework of the notation presented thus far. For a particular property $\mu$, precision is defined as the amount of property values of type $\mu$ received that were relevant relative to the total number of property values retrieved overall. This is represented in Eq. \ref{eq:precision} where the function $\mathrm{rec}(n_i,\mu)$ returns the set of recommended property values for resource $n_i$ of type $\mu$, while $\mathrm{meta}(n_i,\mu)$ returns the set properties values of type $\mu$ previously existing for resource $n_i$. Since the validation is against an artificially atrophied resource set, the recommended property values are checked against the previously existing property values (prior to artificial atrophy).\\

 	\begin{equation*}
		\label{eq:precision}
			Pr(\mu) = \frac{|\mathrm{meta}(n_i,\mu) \cap \mathrm{rec}(n_i,\mu)|}{|\mathrm{rec}(n_i,\mu)|}
	\end{equation*}

Recall, Eq. \ref{eq:recall}, on the other hand, is defined as the proportion of relevant property values received to the total amount of relevant property values possible. For example, if resource $n_i$ previously (before artificial atrophy) had the property value keyword=$\{$swarm$\}$ and is recommended the property value keyword=$\{$swarm$\}$, then there is a 100\% recall.  On the other hand, if resource $n_i$ previously had the property values keyword=$\{$swarm, network$\}$ and is recommended the property value keyword=$\{$swarm$\}$, then there is a 50\% recall, whereas its precision is 100\% in both cases.\\

 	\begin{equation*}
		\label{eq:recall}
			Re(\mu) = \frac{|\mathrm{meta}(n_i,\mu) \cap \mathrm{rec}(n_i,\mu)|}{|\mathrm{meta}(n_i,\mu)|}
	\end{equation*}

Precision and recall tend to be inversely related, $Pr \approx \frac{1}{Re}$. This inverse relationship is understood best when examining the extreme cases. If every possible property value was provided to a resource ($|\mathrm{rec}(n_i,\mu)| \rightarrow \infty$), and that resource originally only had one property value ($|\mathrm{meta}(n_i,\mu)| = 1$) then the recall would be 100\% while the precision would be near 0\%.  At the opposite extreme, if a resource previously had every possible property value in its original metadata ($|\mathrm{meta}(n_i,\mu)| \rightarrow \infty$) and was recommend only one property value ($|\mathrm{rec}(n_i,\mu)| = 1$), then the precision would be 100\%, but the recall would be near 0\%. While, in some systems, precision and recall can be inversely related, it is the goal of information retrieval systems that are validated according to this criterion to achieve both high precision and recall values.\\

Finally, F-score, Eq. \ref{eq:fscore}, can be used to combine precision and recall into a single measure. Note that different associative networks will perform differently for different property types. For instance, co-citation networks will, intuitively, preform better at propagating keyword values than co-organization networks. Therefore, the F-score measure will be represented as $F(\mu_x,\mu_y)$ in order to express the F-score of a network created from metadata properties of type $\mu_y$ propagating property values of type $\mu_x$.  Precision and recall can be represented in a similar fashion though the results of the experiment to follow are expressed according to the F-score measure only.\\

  \begin{equation*}
		\label{eq:fscore}
			F(\mu_x,\mu_y) = \frac{2 \cdot Pr(\mu_x) \cdot Re(\mu_x)}{Pr(\mu_x) + Re(\mu_x)}
	\end{equation*} 

\subsection{Experiment Parameters}

The experiment was set up to determine various F-scores, $F(\mu_x,\mu_y)$, where $\mu_x \in \{\mathrm{auth}, \mathrm{cite}, \mathrm{date}, \mathrm{jour}, \mathrm{key}, \mathrm{org}\}$ and $\mu_y \in \{\mathrm{cite}, \mathrm{coauth}, \mathrm{cocite}, \mathrm{cokey}, \mathrm{coorg}\}$. This means that for every type of associative network generated, an F-score for each metadata property type was determined.  Since the hep-th 2003 bibliographic dataset is a metadata-rich dataset, it was necessary to destroy a percentage of the metadata to test whether or not the metadata generation algorithm could reconstruct the property values for the selected metadata-poor resources. Therefore, the tunable parameter, density, $\partial \in [0.01,0.9]$, was created. The density of the network metadata ranges from 1\% of the network resources containing metadata to 99\% of the resources. Given the percentage parameter, resources were randomly selected for atrophy before the metadata propagation algorithm was run.\\

With the potential for 99\% of the network containing metadata, the propagation of metadata to the lacking 1\% would be overwhelming (a high recall with a low precision). In order to allow nodes to regulate the amount of metadata property values they accept, a second parameter exists. The percentile parameter, $\rho \in [0,1]$, determines the energy threshold for property value recommendations. Since each $\mathrm{rec}(n_i,\mu)$ entry has an associated energy value (recommendation influence), a range from 0$^\mathrm{th}$ percentile, meaning all provided property values are accepted to 100$^\mathrm{th}$ percentile, meaning only the top energy property value is accepted, is used. The pseudo-code for the experimental set-up is presented in Algorithm \ref{alg:experiment}. In Algorithm \ref{alg:experiment}, $\mathrm{killMeta}()$, $\mathrm{acceptMeta}()$, and $\mathrm{calculateF}()$ do not have accompanying pseudo-code.\\

  \restylealgo{boxed}
  \linesnumbered
  \begin{algorithm}[h!]
  \begin{footnotesize}
  \Indp
  	\KwIn{$\mathrm{experiment}()$}
  	\# run the metadata propagation algorithm for each associative network type \;
	  \ForEach{$(\mu_y \in [\mathrm{coauth, cocite, cokey, cite}])$}{
	 \Indp		$\mathrm{loadNetwork}(\mu_y)$ \;
			\ForEach{$(\mu_x \in [\mathrm{auth,cite,date,jour,key,org}])$}{
	 \Indp		\# atrophy a randomly selected percentage of the network \;
			\For{$(\partial = 0.01, \; \partial < 1.0, \; \partial = \partial + 0.2)$}{
	 \Indp			$\mathrm{killMeta}(1 - \partial)$ \;
				$\mathrm{propagateMeta}(\mu_x)$ \;
				\# allow metadata-poor resources to accept only a certain percentage of their recommended property values \;
	 			\For{$(\rho = 0.0, \; \rho <= 1.0, \; \rho = \rho + 0.1)$}{
	 \Indp				$\mathrm{acceptMeta}(\rho)$ \;
					$\mathrm{calcuateF}(\mu_x, \mu_y)$ \;
				 }
				}
			}
	  }
	\caption{\label{alg:experiment} Determining the F-score for the various experimental parameters }
  \end{footnotesize}
  \end{algorithm}

The general expected trend is that as the density of the network increases, the recall increases and the precision decreases. With more property values being propagated, any metadata-poor record, on average, will receive more recommendations than are needed. For instance, a manuscript only has one publishing journal, therefore a recommendation of $100$ journals is going to yield a very low precision ($0.01$).  To balance this effect, a percentile increase will tend to increase the precision of the algorithm at the expense of recall. When only the highest energy recommendations are accepted, the probability of rejecting a useful recommendation increases. In the case of journal propagation, if only the 100$^\mathrm{th}$ percentile recommendation is accepted, then only the highest energy recommendation is accepted.  If this journal recommendation is the correct publishing venue, then there is 100\% recall and precision. If not, then there is 0\% recall and precision. Depending on the amount of values needed to fill a particular property, different $\rho$ values will be most suitable than others.\\

\subsection{The Results}

This section presents the results of the experiment outlined previously in Algorithm \ref{alg:experiment}. For every associative network type and for every metadata type, a F-score matrix was determined for every combination of $\partial$ (density) and $\rho$ (percentile). These F-score values were calculated as the average over $20$ different runs of the experiment. Tables \ref{tab:maxscores} and \ref{tab:meanscores} provide the max and mean F-scores for each network/metadata pair over the entire $\partial$/$\rho$ set. Note that the bold faced values are those $\mu_x$/$\mu_y$ pairs for which a landscape plot is provided. The italicized values are experimental anomalies since the same metadata that was used to generate the associative network is also the same metadata being propagated. For all other combinations, metadata of a particular $\mu$ type exists to create an associative network and metadata properties of a different $\mu$ type is being propagated over those edges. For instance, a co-authorship network is used to propagate citation property values.\\

\begin{table*}[ht!]
	\begin{footnotesize}
	\centering
		\begin{tabular}{|l||c|c|c|c|c|c|}
			\hline
			\textbf{network/metadata} & author & citation & date & journal & keyword & organization \\\hline\hline
			citation        & \textbf{0.1829} & \textit{0.1757} & 0.0606 & \textbf{0.2438} & \textbf{0.3913} & \textbf{0.2782} \\\hline
			co-author       & \textit{0.6218} & \textbf{0.1300} & 0.0717 & \textbf{0.2630} & \textbf{0.2795} & \textbf{0.6457} \\\hline
			co-citation     & 0.0770 & \textit{0.1821} & \textit{0.0780} & \textbf{0.2081} & \textbf{0.2213} & 0.1350 \\\hline
			co-keyword      & 0.0073 & 0.0248 & 0.0472 & \textbf{0.1904} & \textit{0.8689} & 0.0420 \\\hline
			co-organization & 0.0709 & 0.0236 & 0.0508 & \textbf{0.1918} & 0.1180 & \textit{0.5000} \\\hline
		\end{tabular}
	\caption{\label{tab:maxscores} Max F-scores}
	\end{footnotesize}
\end{table*}

\begin{table*}[ht!]
  \begin{footnotesize}
	\centering
		\begin{tabular}{|l||c|c|c|c|c|c|}
			\hline
			\textbf{network/metadata} & author & citation & date & journal & keyword & organization \\\hline\hline
			citation        & \textbf{0.1367} & \textit{0.1327} & 0.0441 & \textbf{0.2133} & \textbf{0.3246} & \textbf{0.2218} \\\hline
			co-author       & \textit{0.2848} & \textbf{0.0780} & 0.0548 & \textbf{0.2004} & \textbf{0.1958} & \textbf{0.3935} \\\hline
			co-citation     & 0.0338 & \textit{0.0697} & 0.0539 & \textbf{0.1554} & \textbf{0.1509} & 0.0768 \\\hline
			co-keyword      & 0.0032 & 0.0160 & 0.0385 & \textbf{0.1468} & \textit{0.3240} & 0.0330 \\\hline
			co-organization & 0.0312 & 0.0145 & 0.0392 & \textbf{0.1410} & 0.0909 & \textit{0.1554} \\\hline
		\end{tabular}
	\caption{\label{tab:meanscores} Mean F-scores}
	\end{footnotesize}
\end{table*}

The following landscape plots expose the relationship between $\partial$ and $\rho$. A short explanation of the intuition behind each plot is also provided.\\

\begin{figure*}[ht!]
	\centering
		\scalebox{0.65}[0.65]{
		  \rotatebox{0}{\includegraphics[width=0.50\textwidth]{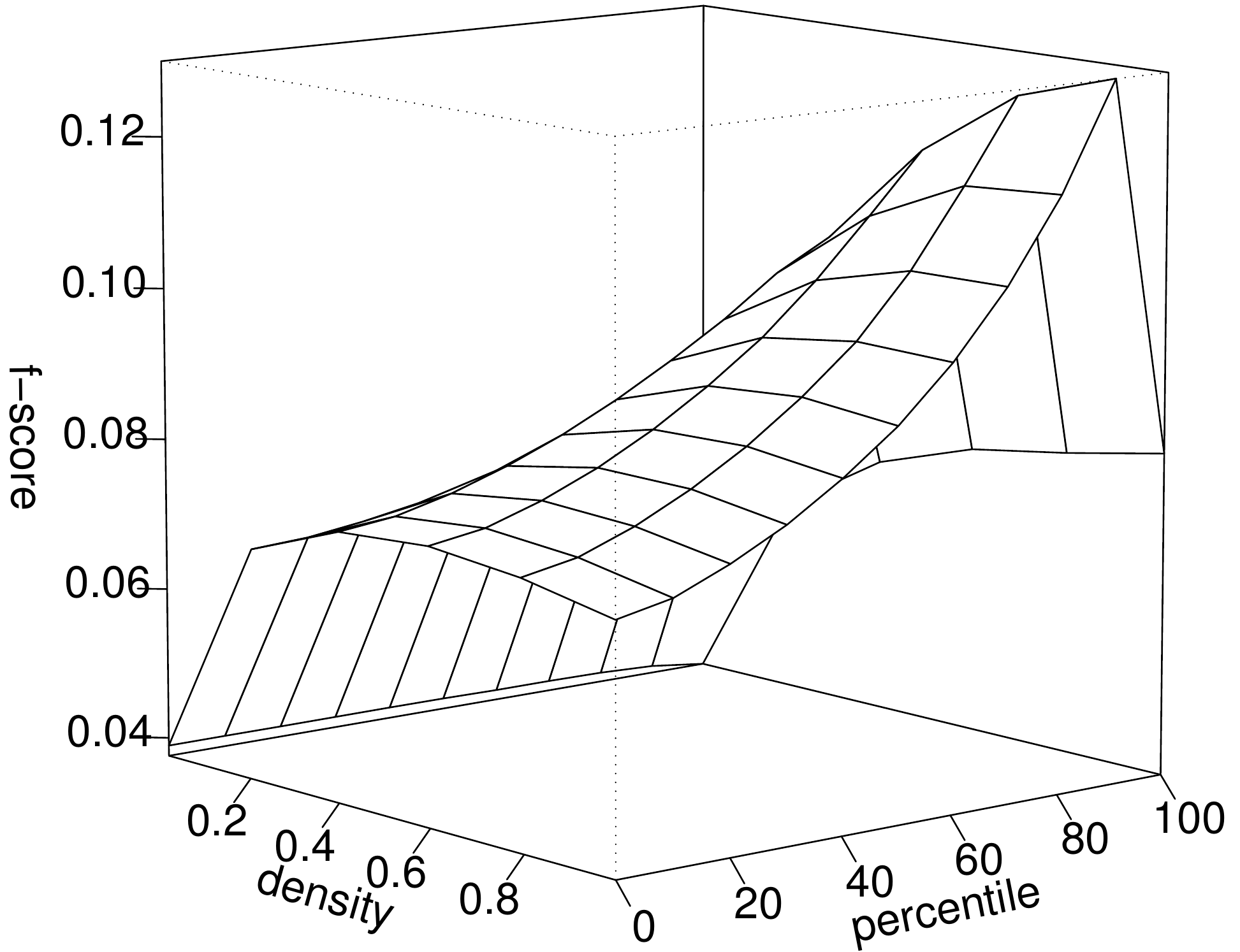}}
		  \rotatebox{0}{\includegraphics[width=0.50\textwidth]{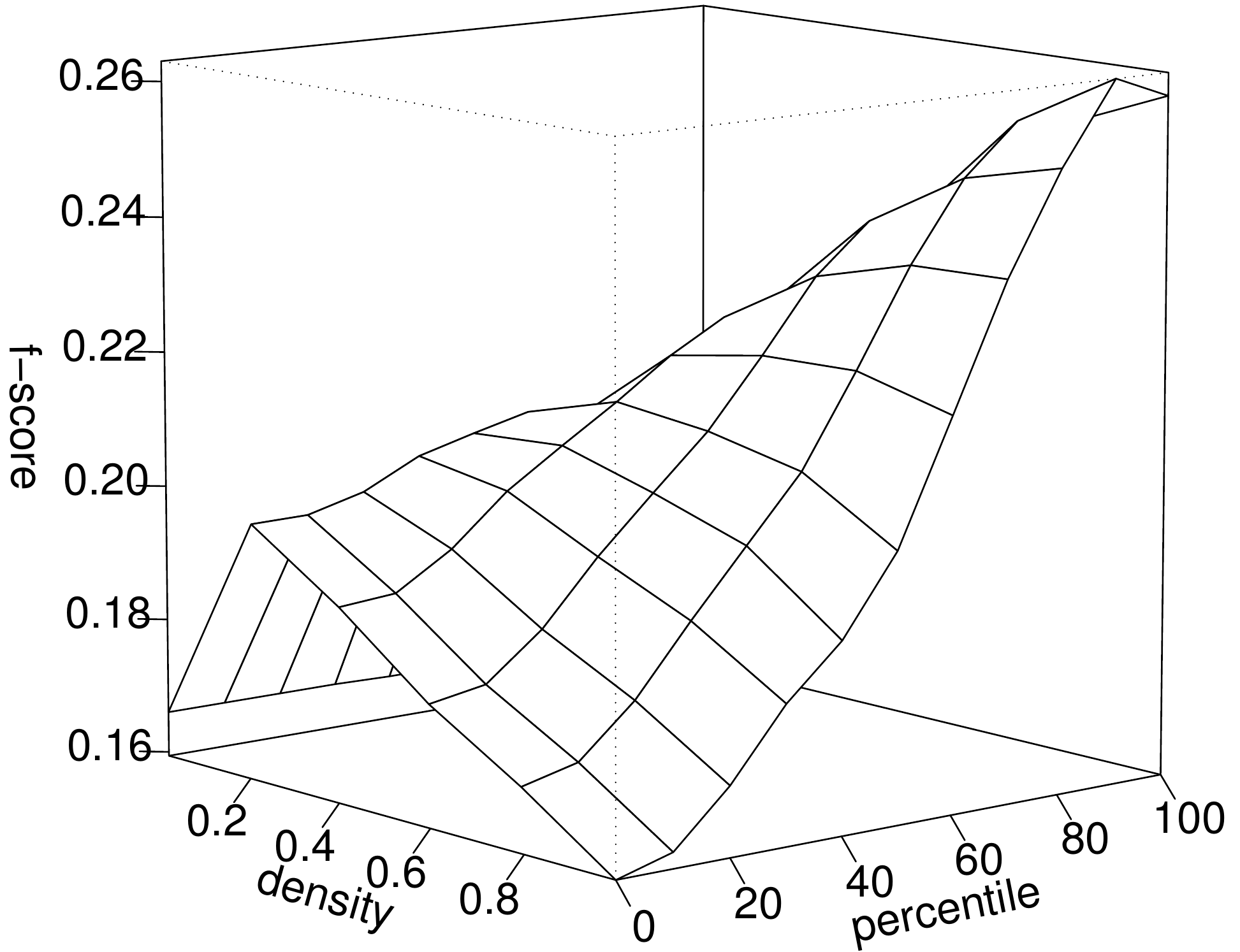}}}
	\caption{Co-authorship network propagating \textbf{a.} citation $F(\mathrm{cite},\mathrm{coauth})$ and \textbf{b.} journal $F(\mathrm{jour},\mathrm{coauth})$ metadata}
	\label{fig:author1}
\end{figure*}

\begin{figure*}[ht!]
	\centering
		\scalebox{0.65}[0.65]{
		  \rotatebox{0}{\includegraphics[width=0.50\textwidth]{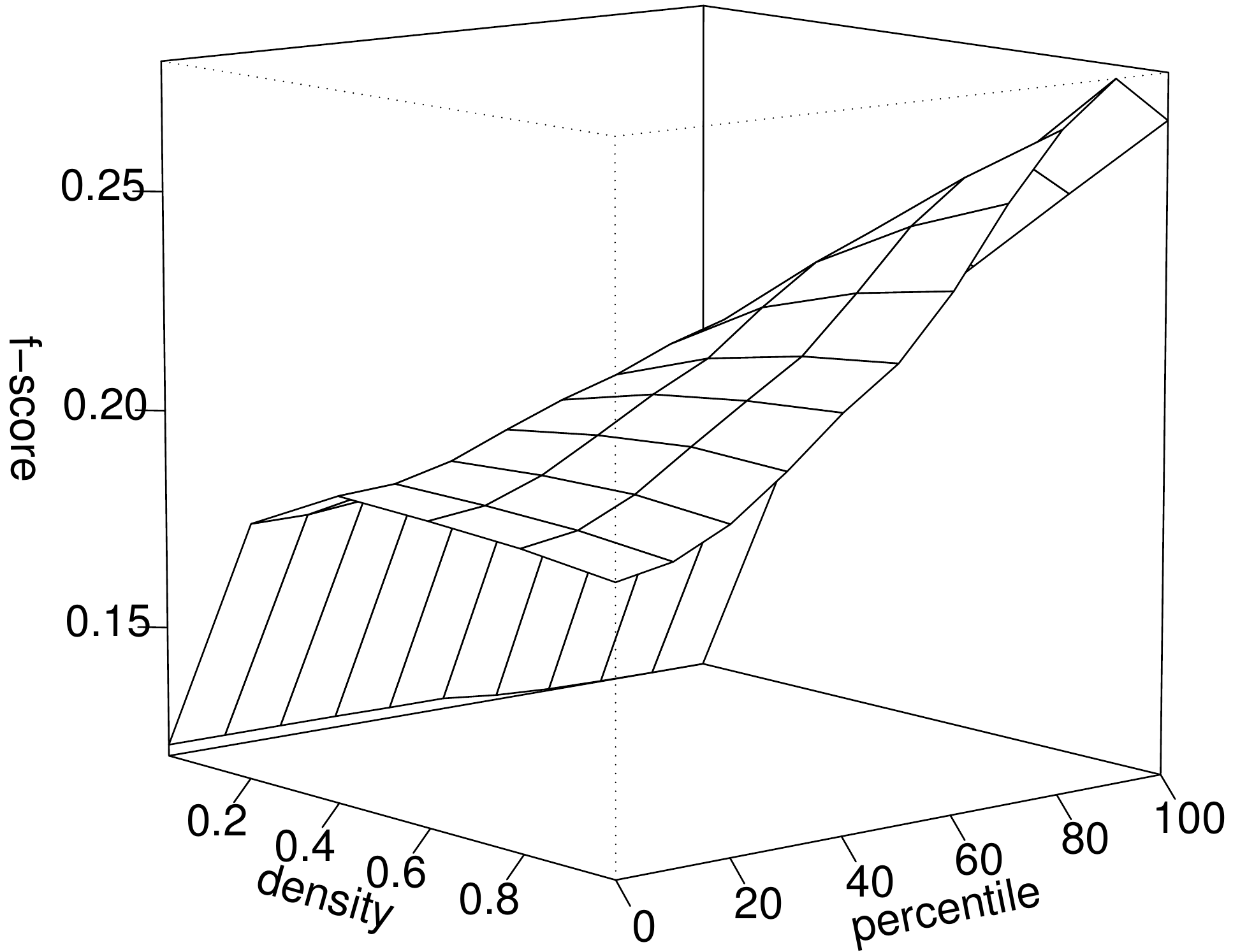}}
		  \rotatebox{0}{\includegraphics[width=0.50\textwidth]{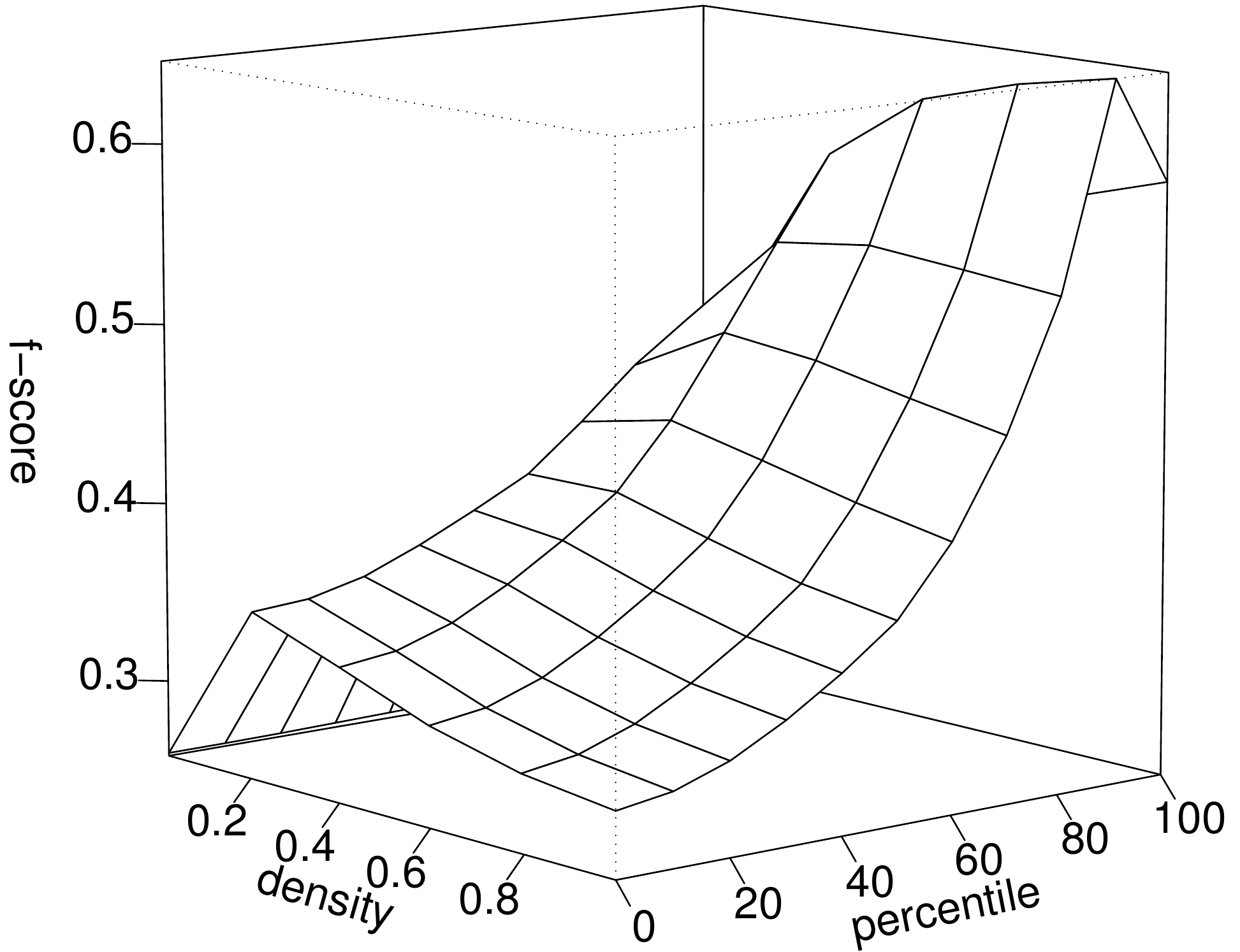}}}
	\caption{Co-authorship network propagating \textbf{a.} keyword $F(\mathrm{key},\mathrm{coauth})$ and \textbf{b.} organization $F(\mathrm{org},\mathrm{coauth})$ properties}
	\label{fig:author2}
\end{figure*}

Intuitively, it makes sense that a co-authorship network would perform well when propagating citation, journal, keyword, and organization property values which are represented in Figure \ref{fig:author1}a, Figure \ref{fig:author1}b, Figure \ref{fig:author2}a, Figure \ref{fig:author2}b respectively. The performance is a result of the fact that collaborating authors tend to cite themselves, publish in similar journals, write about similar topics, and are within similar organizations. Notice the effect that percentile ($\rho$) has on Figure \ref{fig:author1}a as opposed to Figure \ref{fig:author1}b. Since there tend to exist many citation property values (manuscripts cite many manuscripts), lower percentile values ($\rho \approx 0$) ensures that there is a high recall. When $\rho = 1.0$, only the top citation is accepted and therefore the F-score drops (very poor recall). On the other other hand, in Figure \ref{fig:author2}, when $\rho = 0.0$, there are many journal recommendations. This is not desirable since a journal property only has one value (a manuscript is published in only one venue). Therefore, at $\rho = 1.0$, only one journal value is accepted into the resource's journal property. In situations where few property values are expected, the F-score is best with a high $\rho$.\\

\begin{figure*}[ht!]
	\centering
		\scalebox{0.65}[0.65]{
		  \rotatebox{0}{\includegraphics[width=0.50\textwidth]{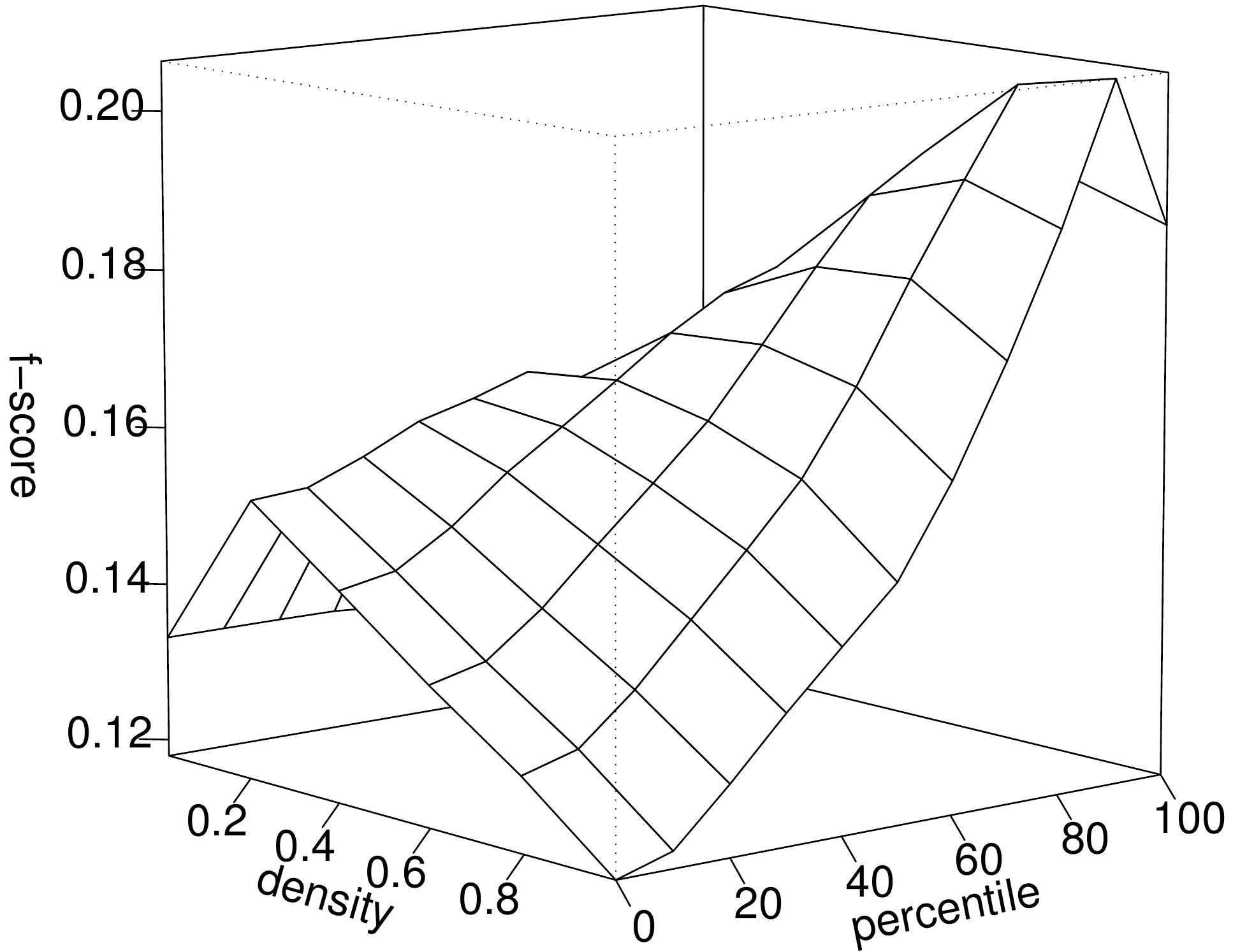}}
		  \rotatebox{0}{\includegraphics[width=0.50\textwidth]{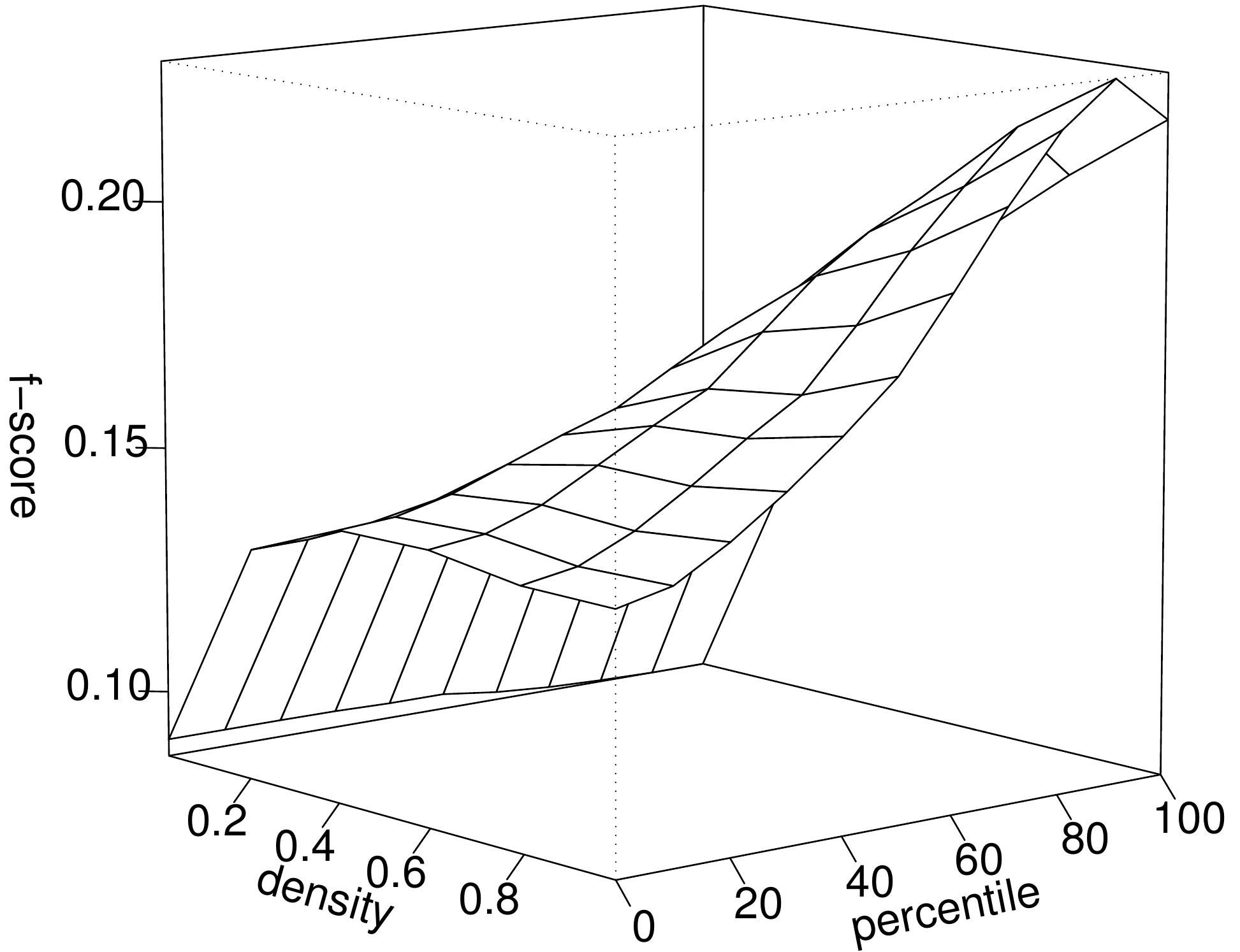}}}
	\caption{Co-citation network propagating \textbf{a.} journal $F(\mathrm{jour},\mathrm{cocite})$ and \textbf{b.} keyword $F(\mathrm{key},\mathrm{cocite})$ properties}
	\label{fig:cocite1}
\end{figure*}

A co-citation network, Figure \ref{fig:cocite1}, performs best with journal and keyword properties.  This means that manuscripts are likely to cite other manuscripts with similar journal venues and since citation tends to be within the same subject domain, the probability of similar keyword metadata increases. Again, notice the effect of $\rho$ on journal metadata propagation. The shape of the Figure \ref{fig:cocite1}a graph nearly mimics the shape of Figure \ref{fig:author1}b. Likewise, for Figure \ref{fig:cocite1}b and Figure \ref{fig:author2}a. Again, the expected property value number is a major factor in determining the system's $\rho$ parameter.\\

\begin{figure*}[ht!]
	\centering
		\scalebox{0.65}[0.65]{
		  \rotatebox{0}{\includegraphics[width=0.50\textwidth]{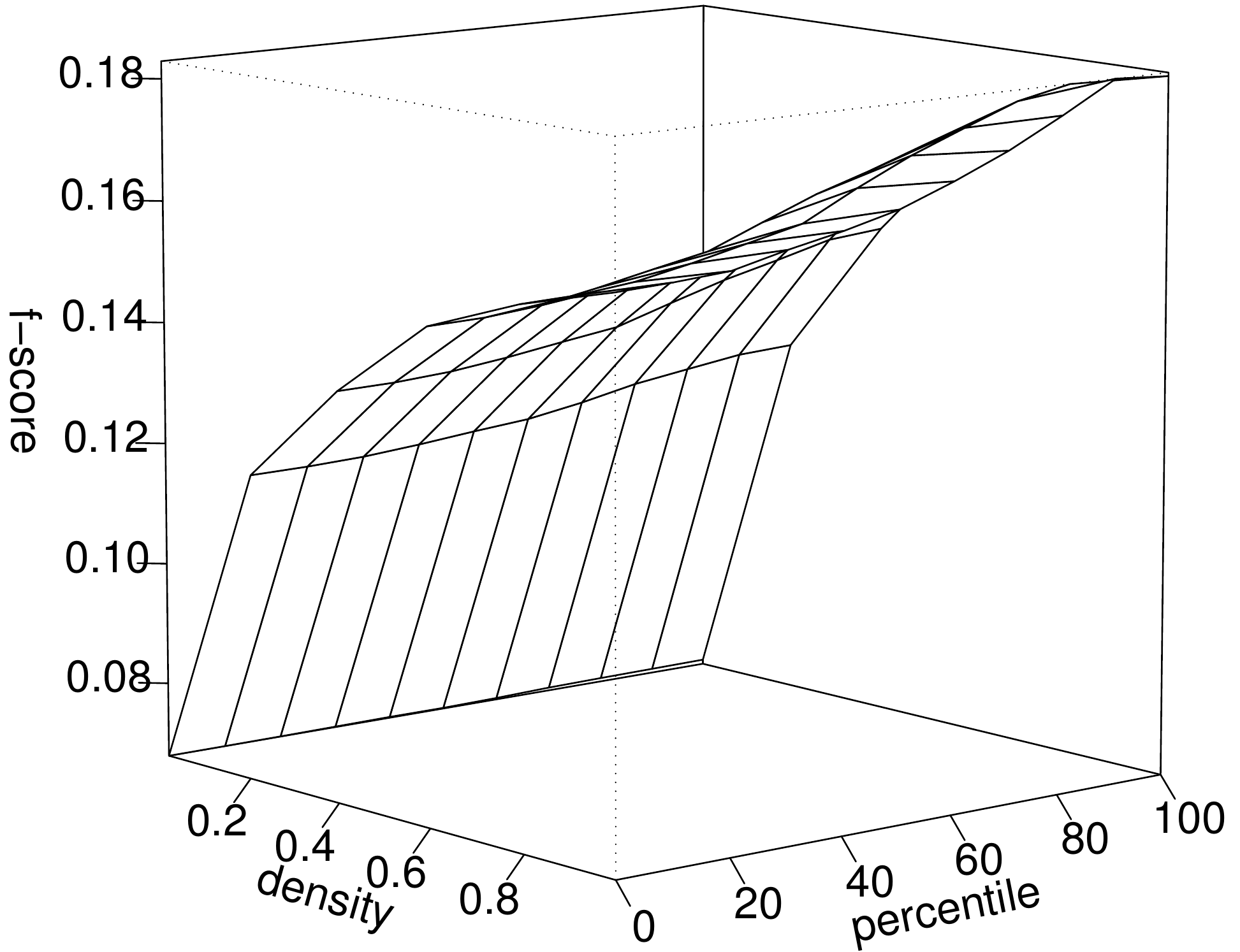}}
		  \rotatebox{0}{\includegraphics[width=0.50\textwidth]{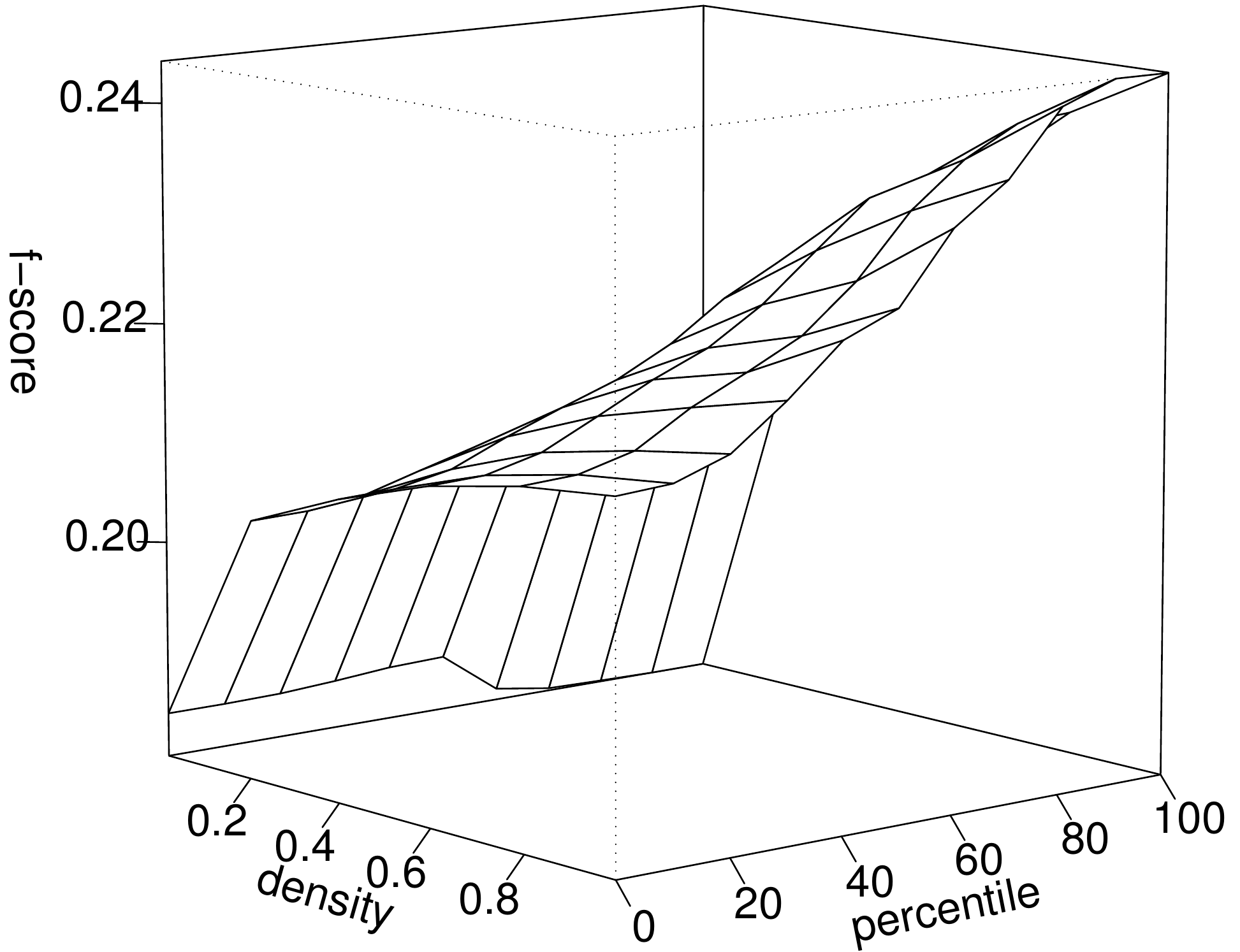}}}
	 \caption{Citation network propagating \textbf{a.} author $F(\mathrm{auth},\mathrm{cite})$ and \textbf{b.} journal $F(\mathrm{jour},\mathrm{cite})$ properties}
	\label{fig:cite1}
\end{figure*}

\begin{figure}[h!]
	\centering
		\includegraphics[width=0.25\textwidth]{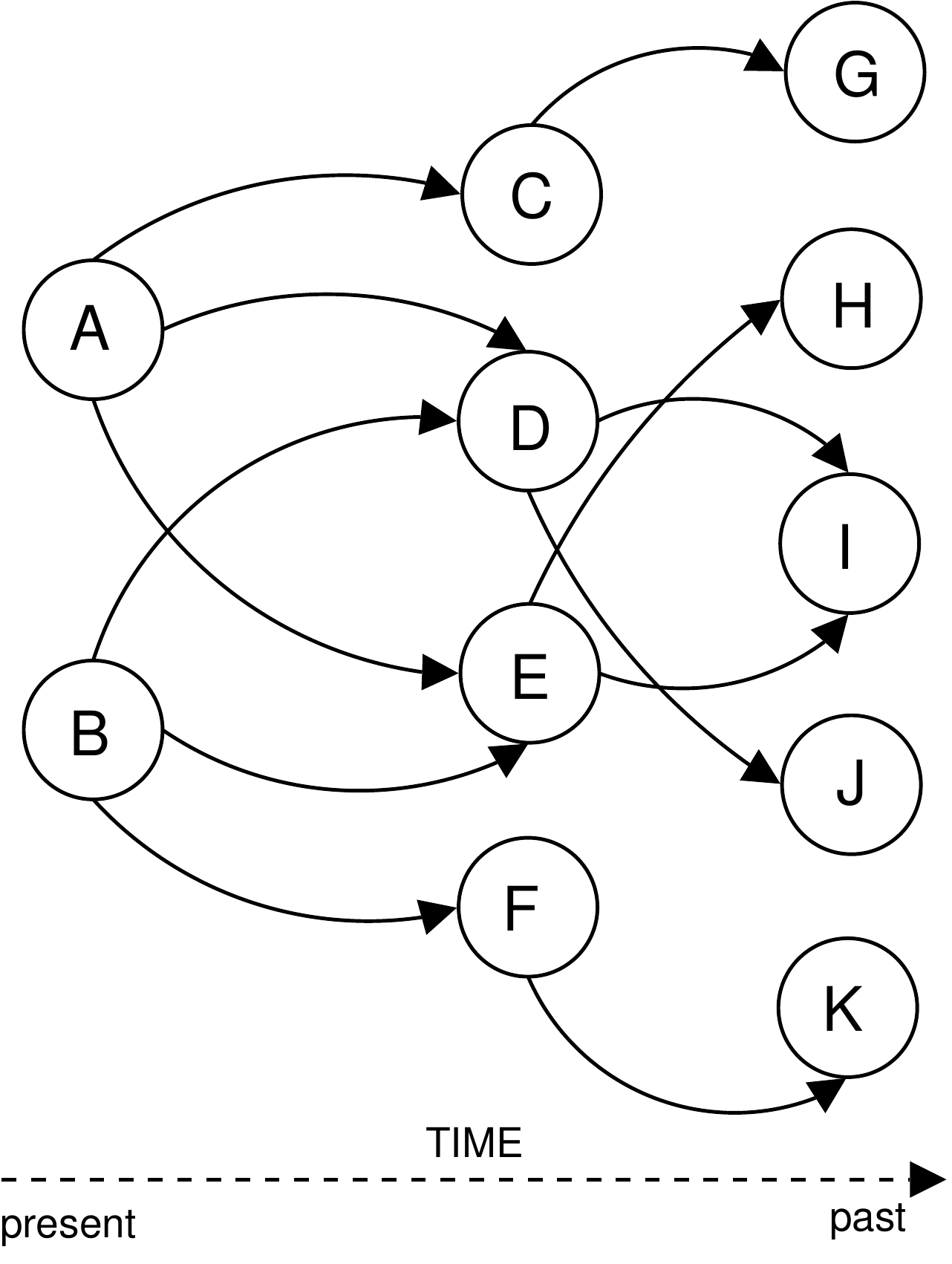}
	\caption{Citation networks are non-recurrent networks}
	\label{fig:citation}
\end{figure}

\begin{figure*}[ht!]
	\centering
		\scalebox{0.65}[0.65]{
		  \rotatebox{0}{\includegraphics[width=0.50\textwidth]{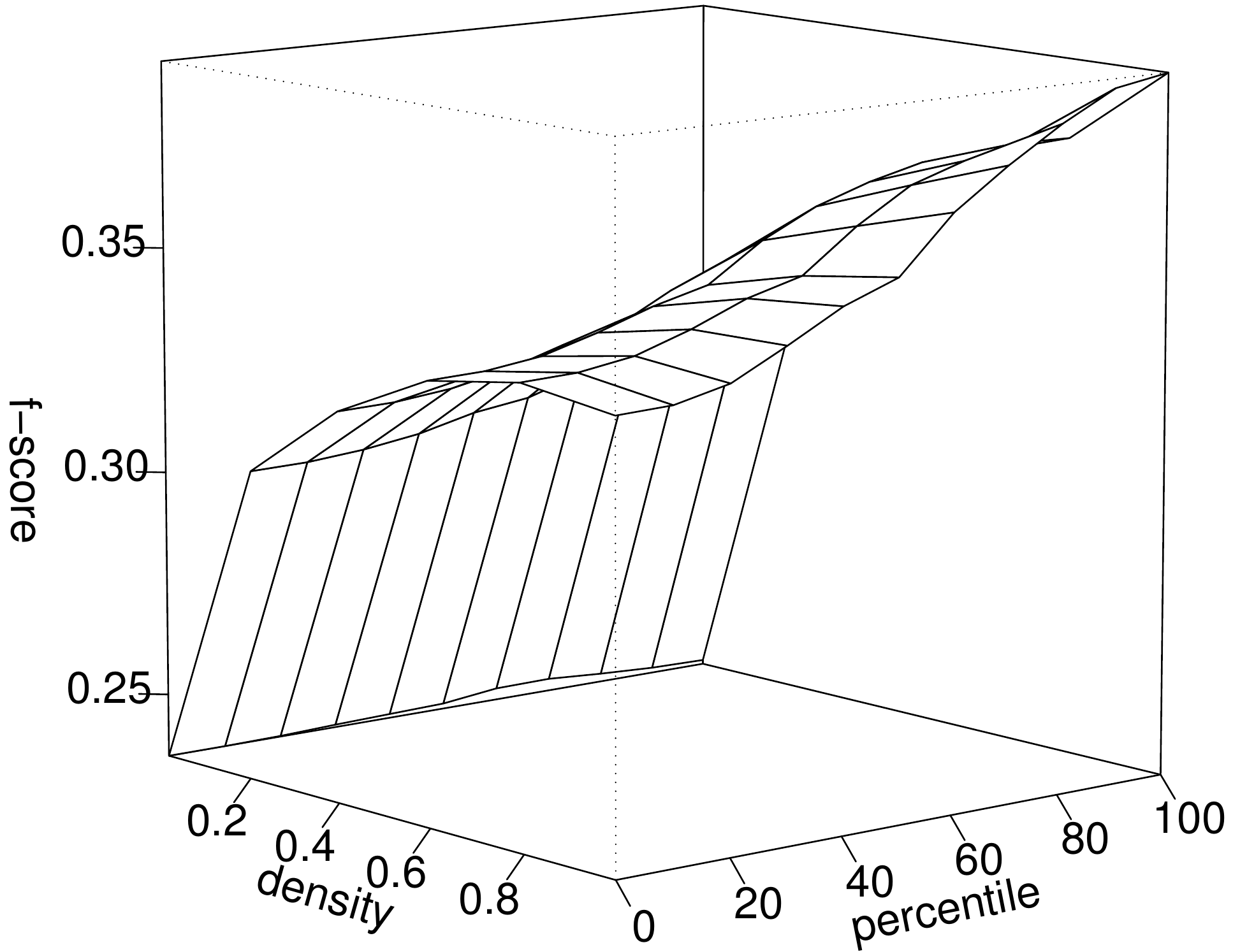}}
		  \rotatebox{0}{\includegraphics[width=0.50\textwidth]{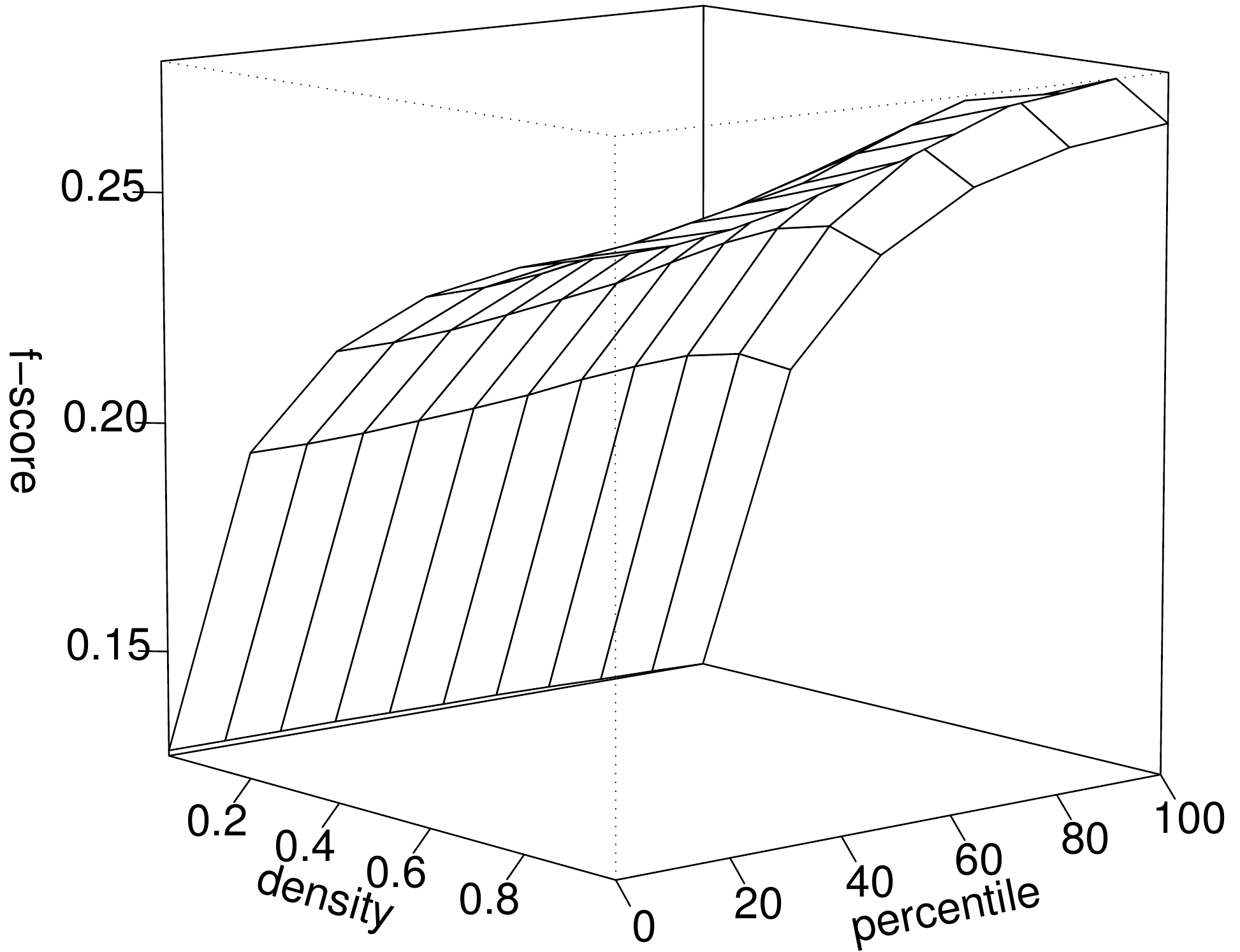}}}
	\caption{Citation network propagating \textbf{a.} keyword $F(\mathrm{key},\mathrm{cite})$ and \textbf{b.} organization $F(\mathrm{org},\mathrm{cite})$ properties}
	\label{fig:cite2}
\end{figure*}

A citation network, like a co-citation network performs well for author, journal, keyword, and organizational properties Figure \ref{fig:cite1} and Figure \ref{fig:cite2}. It is interesting to note how much better a citation network works for $\rho \approx 0.0$. Since a citation network isn't symmetric, there is a chance that a particle will reach a dead end. When a particle reaches a dead end, it no longer recommends property values. Furthermore, citations are in a hierarchy with more recent publications being at the top of the hierarchy (manuscripts can not cite forward in time). Particles therefore trickle down the hierarchy via a single, non-recurrent path from top to bottom. This ``plinko ball" effect is represented in Figure \ref{fig:citation}. The lack of recurrence in citation networks tends to produce a high precision with a lower recall.  High precision and low recall is exactly what a low $\rho$ produces. Therefore, since the topology of the citation network yields the same effect, the effect of $\rho$ as $\rho \rightarrow 0.0$ isn't as pronounced.\\ 

As can be noticed from Table \ref{tab:maxscores}, Table \ref{tab:meanscores}, and Figure \ref{fig:cite2}a, the keyword property performs best in a citation network. A direct reference from one document to another is a validation of the similarity between documents with respect to subject domain.  Therefore, the tendency for citing documents to contains similar keyword values is high. For instance, refer to the citations of this article (references in this manuscript's bibliography). Every cited manuscript is either about automatic metadata generation, bibliographic networks, or network analysis.\\

\begin{figure*}[ht!]
	\centering
		  \scalebox{0.65}[0.65]{
		  	\rotatebox{0}{\includegraphics[width=0.50\textwidth]{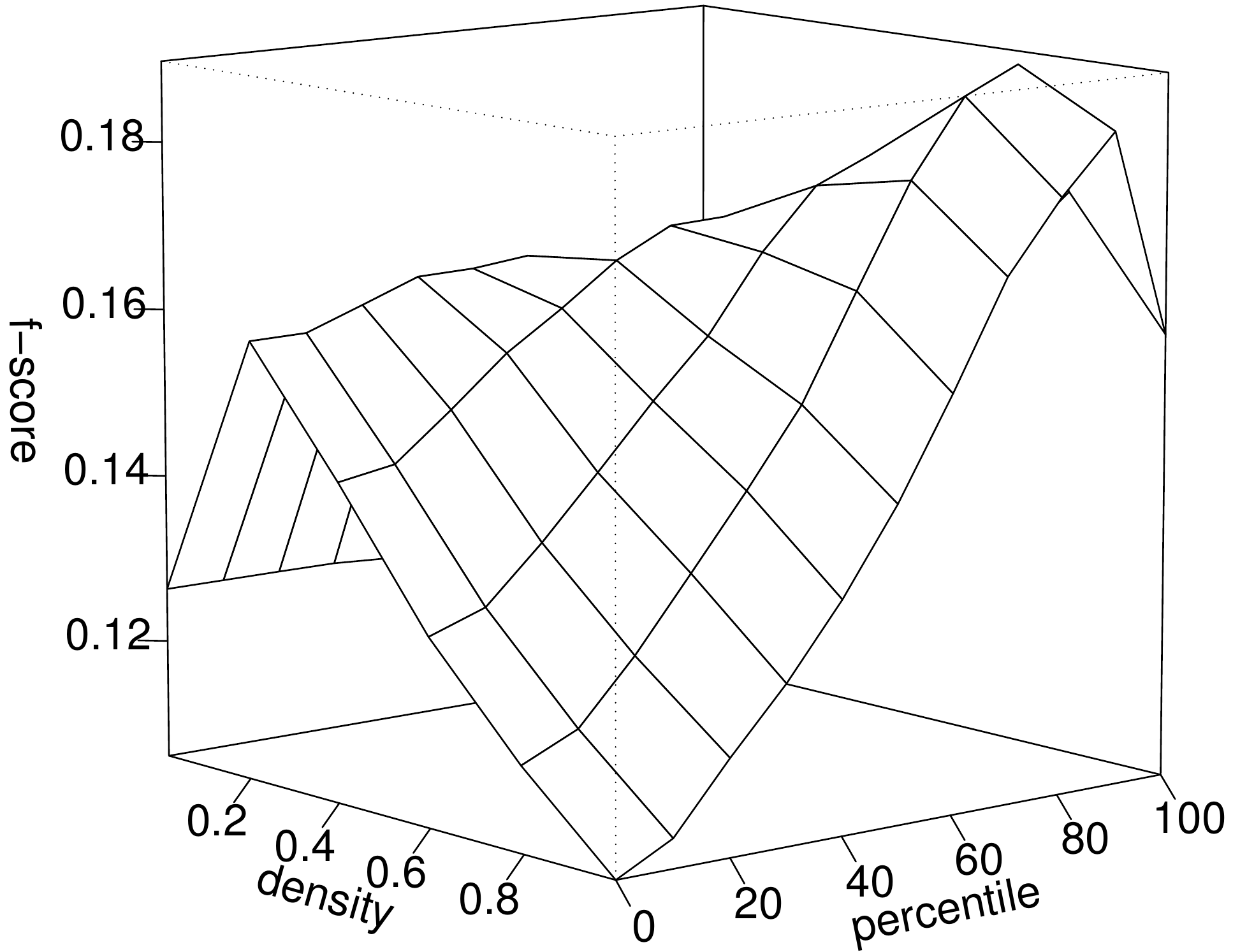}}
		  	\rotatebox{0}{\includegraphics[width=0.50\textwidth]{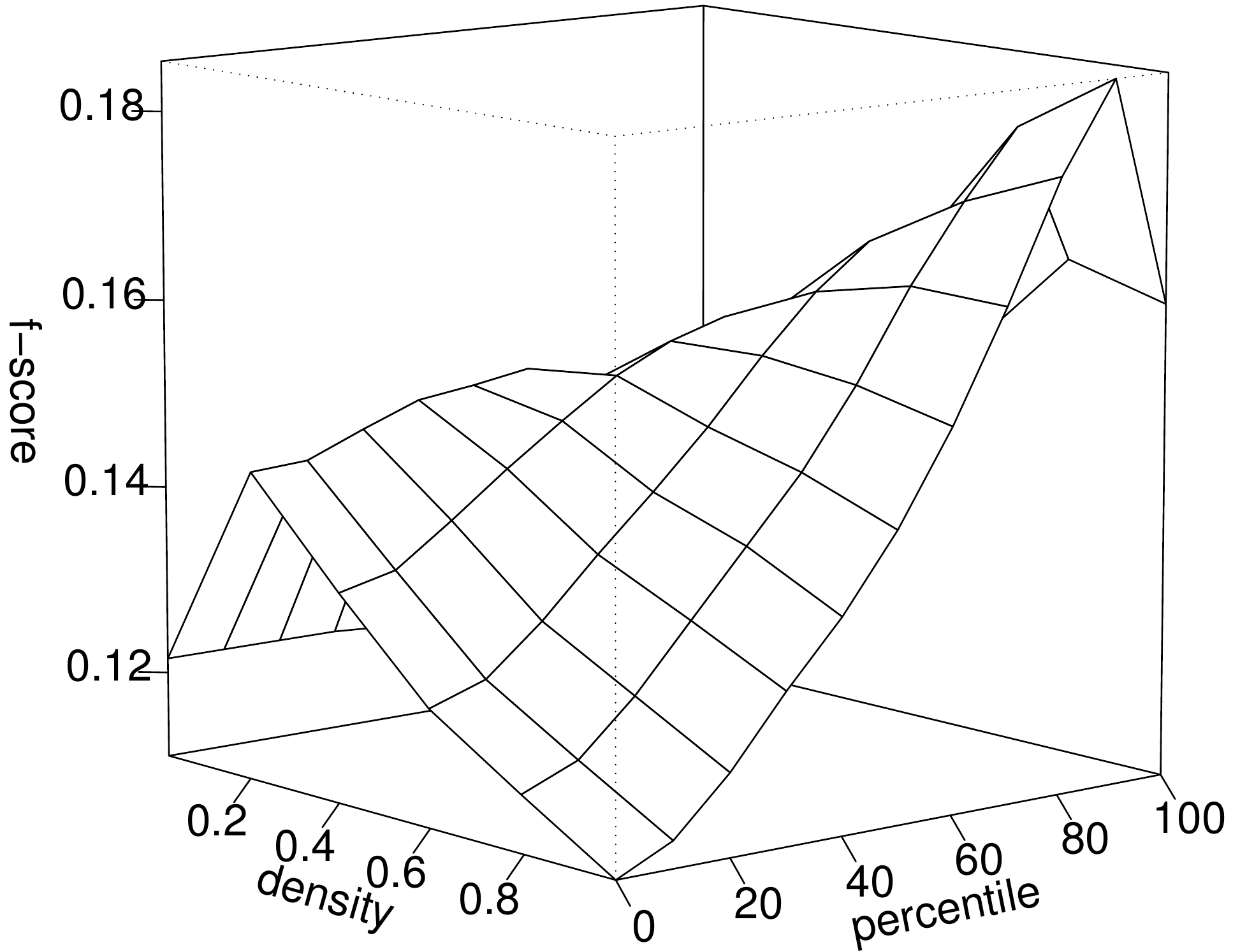}}}
	\caption{Co-keyword and co-organization networks propagating \textbf{a.} journal $F(\mathrm{jour},\mathrm{cokey})$ and \textbf{b.} $F(\mathrm{jour},\mathrm{coorg})$ properties, respectively}
	\label{fig:keyorg}
\end{figure*}

A co-keyword network does not perform well for most property types except the journal property, Figure \ref{fig:keyorg}a.  This makes sense since manuscripts on similar topics are likely to be published in similar journals.\\

\section{Future Work}

This paper has provided a preliminary exploration of metadata generation in terms of metadata property propagation within an associative network of repository resources. Further research in this area may prove useful for other network types such as those generated from other metadata properties.  For instance, it may be of interest to study the effect of this algorithm on usage networks \cite{altern:bollen2005}. Usage metadata, unlike citation and journal metadata, is applicable to every accessible resource. It would be interesting to see what co-usage means for a particular genera of resources by determining which metadata properties these networks are best at propagating.\\

A variety of propagation algorithms may also be explored.  It is assumed that a particle will take only edges of a particular $\mu$ type for the duration of their life-span.  Different path types might be an important aspect of increasing the precision and recall performance of this method.  For instance, keyword metadata that first propagates over co-authorship edges and then over co-citation edges might provide better results. Methods to implement such propagation algorithms have been presented in \cite{socialgrammar:rodriguez2007,grammar:rodriguez2007}.  Also, different edge types can be merged such that all co-keyword and co-authorship edges are collapsed to form a single edge.\\ 

What has been presented in this study is the results of this algorithm without the intervention of any human components (besides the initial creation of metadata through the hep-th dataset creation process). Future work that studies this method with the inclusion of humans that help to validate and ``clean" the recommended metadata would be telling of how much this method is able to speed up the process of generating accurate and reliable metadata for metadata-poor resources. Such an analysis is left to future research.\\

Finally, multiplicative effects due to particle interaction may effect the results of the algorithm. For instance, if two particles, $p_i$ and $p_j$, meet at a particular node, $n_k$, and $p_i$ and $p_j$ have similar metadata then the footprint they leave at $n_k$ should be more noticeable. Because two different metadata sources are supplying the same property values, there is an increased probability of that recommended metadata value being correct. Currently, only a summation is being provided. It may be interesting to multiply this summation by the number of unique particles that provided energy for a particular recommended metadata value. The variations of this preliminary framework will be explored in future work.

\section{Conclusion}

Automatic metadata generation is becoming an increasingly important field of research as digital library repositories become more prevalent and move into the arena of less strongly controlled, decentralized collections (e.g.~arXiv and CiteSeer). The creation and maintenance of high-quality, detailed metadata is hampered on numerous levels. Manual metadata creation methods are costly. Recent efforts to leverage the collective power of social tagging (i.e.~``folksonomies") may address some of the shortcomings of the manual creation of metadata and result in viable models for online resources that do not require strongly controlled vocabularies and metadata ontologies. However, it is doubtful that ``folksonomies" can be generalized to situations that require vetted, well-standardized metadata. The automated creation of metadata on the basis of content-analysis is a promising alternative to the manual creation of metadata. It is conceivably more efficient in situations where textual data is available and allows for more formal control of the type and nature of metadata that is extracted. However, it can be unreliable for non-text resources, yield low-quality metadata and can be computationally expensive.\\

This article proposed another possible component of the metadata generation toolkit which may complement and support the above mentioned approaches. Instead of creating new metadata, metadata is propagated from a metadata-rich subset of the collection to similar, but metadata-poor subsets. The substrate for this extrapolation is an associative network of resource relations created from other available metadata.  Metadata propagation may provide a computationally feasible means of generating large amounts of metadata for heterogeneous resources which can later be fine-tuned by manual intervention or cross-validation with content-based methods. The article finally provided experimental results using the High-Energy Physics bibliographic data set (hep-th 2003). Human intervention may play an important role in fine-tuning the metadata propagation algorithm. The results of this experiment are promising and there still exists a range of potential modifications to this basic framework that may lead to even better results.

\section{Acknowledgements}

This research was financially supported by the Research Library at the Los Alamos National Laboratory. The modified hep-th 2003 bibliographic dataset was generously provided by Shou-de Lin and Jennifer H. Watkins provided editorial assistance. Finally,  the hep-th 2003 database is based on data from the arXiv archive and the Stanford Linear Accelerator Center SPIRES-HEP database provided for the 2003 KDD Cup competition with additional preparation performed by the Knowledge Discovery Laboratory, University of Massachusetts Amherst.\\


\begin{thebibliography}{10}

\bibitem{altern:bollen2005}
Johan Bollen, Herbert {Van de Sompel}, Joan Smith, and Rick Luce.
\newblock Toward alternative metrics of journal impact: a comparison of
  download and citation data.
\newblock {\em Information Processing and Management}, 41(6):1419--1440, 2005.

\bibitem{inform:cohen1987}
Paul~R. Cohen and Rick Kjeldsen.
\newblock Information retrieval by constrained spreading activation in semantic
  networks.
\newblock {\em Information Processing and Management}, 23(4):255--268, 1987.

\bibitem{spread:collins1975}
A.M. Collins and E.F. Loftus.
\newblock A spreading activation theory of semantic processing.
\newblock {\em Psychological Review}, 82:407--428, 1975.

\bibitem{applic:crestani1997}
Fabio Crestani.
\newblock Application of spreading activation techniques in information
  retrieval.
\newblock {\em Artificial Intelligence Review}, 11(6):453--582, 1997.

\bibitem{search:crestani2000}
Fabio Crestani and Puay~Leng Lee.
\newblock Searching the web by constrained spreading activation.
\newblock {\em Information Processing and Management}, 36(4):585--605, 2000.

\bibitem{duval:meta2002}
Erik Duval, Wayne Hodgins, Stuart Sutton, and Stuart~L. Weibel.
\newblock Metadata principles and practices.
\newblock {\em D-Lib Magazine}, 8(4), 2002.

\bibitem{meta:guirida2000}
G.~Giurida and J.~Shek, E.~Yang.
\newblock Knowledge-based metadata extraction from postscript file.
\newblock In {\em {P}roceedings of the {I}nternational {C}onference on
  {D}igital {L}ibraries}, pages 77--84. {ACM}, May 2000.

\bibitem{tagging:hub2006}
Scott~A. Golder and Bernardo~A. Huberman.
\newblock Usage patterns of collaborative tagging systems.
\newblock {\em Journal of Information Science}, 32(2):198--208, 2006.

\bibitem{meta:greenburg2004}
Jane Greenburg.
\newblock Metadata extraction and harvesting: A comparison of two automatic
  metadata generation applications.
\newblock {\em Journal of Internet Cataloging}, 6(4):59--82, 2004.

\bibitem{meta:han2003}
Hui Han, C.~Lee Giles, Eren Manavoglu, and Hongyuan Zha.
\newblock Automatic document metadata extraction using support vector machines.
\newblock In {\em {P}roceedings of the {J}oint {C}onference on {D}igital
  {L}ibraries ({JCDL03})}, Huston, TX, May 2003. {ACM}.

\bibitem{collab:herlocker2006}
Johnathan~L. Herlocker, Joseph~A. Konstan, Loren~G. Terveen, and John~T. Riedl.
\newblock Evaluating collaborative filtering recommender systems.
\newblock {\em ACM Transactions on Information Systems}, 22(1):5--53, 2004.

\bibitem{applyi:huang2004}
Zan Huang, Hsinchun Chen, and Daniel Zeng.
\newblock Applying associative retrieval techniques to alleviate the sparsity
  problem in collaborative filtering.
\newblock {\em ACM Trans. Inf. Syst.}, 22(1):116--142, 2004.

\bibitem{linkpr:hunag2005}
Zan Huang, Xin Li, and Hsinchun Chen.
\newblock Link prediction approach to collaborative filtering.
\newblock In {\em {P}roceedings of the {J}oint {C}onference on {D}igital
  {L}ibraries ({JCDL05})}, Denver, CO, June 2005. {ACM}.

\bibitem{kuwano:meta2004}
H.~Kuwano, Y.~Matsuo, and K.~Kawazoe.
\newblock Reducing the cost of metadata generation by using video/audio
  indexing and natural language processing techniques.
\newblock {\em {NTT} Technical Review}, pages 68--74, 2004.

\bibitem{linkpr:liben2003}
David Liben-Nowell and Jon Kleinberg.
\newblock The link prediction problem for social networks.
\newblock In {\em {T}welfth {I}nternational {C}onference on {I}nformation and
  {K}nowledge {M}anagement}, pages 556--559. {ACM}, November 2003.

\bibitem{unsupervised:lin2004}
{Shou-de} Lin and Hans Chalupsky.
\newblock Issues of verification for unsupervised discovery systems.
\newblock In {\em {K}{D}{D}04 {W}orkshop on {L}ink {D}iscovery}, Seattle, WA,
  2004.

\bibitem{meta:mao2004}
Song Mao, Jong~Woo Kim, and George~R. Thoma.
\newblock A dynamic feature generation system for automated metadata extraction
  in preservation of digital materials.
\newblock In {\em {F}irst {I}nternational {W}orkshop on {D}ocument {I}mage
  {A}nalysis for {L}ibraries {DIAL}'04}. {IEEE}, January 2004.

\bibitem{tag:mathes2004}
Adam Mathes.
\newblock Folksonomies - cooperative classification and communication through
  shared metadata.
\newblock Computer Mediated Communication - LIS590CMC (graduate course),
  December 2004.

\bibitem{mcgovern03exploiting}
A.~McGovern, L.~Friedland, M.~Hay, B.~Gallagher, A.~Fast, J.~Neville, and
  D.~Jensen.
\newblock Exploiting relational structure to understand publication patterns in
  high-energy physics.
\newblock {\em {ACM SIGKDD} Explorations Newsletter}, 5(2):165--172, December
  2003.

\bibitem{levera:naaman2005}
Mor Naaman, Ron~B. Yeh, Hector Garcia-Molina, and Andreas Paepcke.
\newblock Leveraging context to resolve identity in photo albums.
\newblock In {\em {P}roceedings of the 5th {J}oint {C}onference on {D}igital
  {L}ibraries ({JCDL} 05)}, Denver, {CO}, June 2005.

\bibitem{metaprop:prime2005}
Camille Prime-Claverie, Michel Beigbeder, and Thierry Lafouge.
\newblock Metadata propagation in the web using co-citations.
\newblock In {\em {P}roceedings of the 2005 {IEEE}/{ACM} {I}nternational
  {C}onference on {W}eb {I}ntelligence}, Compiegne University of Technology,
  France, September 2005. {IEEE} Computer Society.

\bibitem{socialgrammar:rodriguez2007}
Marko~A. Rodriguez.
\newblock Social decision making with multi-relational networks and
  grammar-based particle swarms.
\newblock In {\em {P}roceedings of the {H}awaii {I}nternational {C}onference on
  {S}ystems {S}cience}, pages 39--49, Waikoloa, Hawaii, January 2007. {IEEE}
  Computer Society.

\bibitem{grammar:rodriguez2007}
Marko~A. Rodriguez.
\newblock Grammar-based random walkers in semantic networks.
\newblock {\em Knowledge-Based Systems}, 21(7):727--739, 2008.

\bibitem{sebastiani02machine}
Fabrizio Sebastiani.
\newblock Machine learning in automated text categorization.
\newblock {\em ACM Computing Surveys}, 34(1):1--47, 2002.

\bibitem{dc:ward2003}
Jewel Ward.
\newblock A quantitative analysis of unqualified dublin core metadata element
  set usage within data providers registered with the open archives initiative.
\newblock In {\em JCDL '03: Proceedings of the 3rd ACM/IEEE-CS joint conference
  on Digital libraries}, pages 315--317, Washington, DC, USA, 2003. IEEE
  Computer Society.

\bibitem{meta:yang2005}
Hsin-Chang Yang and Chung-Hong Lee.
\newblock Automatic metadata generation for web pages using a text mining
  approach.
\newblock In {\em {I}nternational {W}orkshop on {C}hallenges in {W}eb
  {I}nformation {R}etrieval and {I}ntegration}, pages 186--194. {IEEE}, April
  2005.

\end{thebibliography}
\end{document}